%% file: main.tex
\documentclass[journal]{IEEEtran}
\IEEEoverridecommandlockouts

\usepackage{cite}
\usepackage{amsmath,amssymb,amsfonts}
\usepackage{algorithmic}
\usepackage{graphicx}
\usepackage{textcomp}
\usepackage{xcolor}
\usepackage{graphicx}
\usepackage{array}
\usepackage{caption}
\usepackage{booktabs}
\usepackage{multirow,makecell}
\usepackage{soul}
\usepackage{subcaption}
\usepackage{balance}
\usepackage[capitalise]{cleveref}
\crefname{section}{Sec.}{Secs.}
\crefname{figure}{Fig.}{Figs.}
\usepackage{tikz}
\usetikzlibrary{fit}
\makeatletter
\tikzset{
  fitting node/.style={
    inner sep=0pt,
    fill=none,
    draw=none,
    reset transform,
    fit={(\pgf@pathminx,\pgf@pathminy) (\pgf@pathmaxx,\pgf@pathmaxy)}
  },
  reset transform/.code={\pgftransformreset}
}
\makeatother

\usepackage{pgfplots}
\pgfplotsset{compat=newest} 
\pgfplotsset{plot coordinates/math parser=false} 

\usepackage[acronyms,nonumberlist,nopostdot,nomain,nogroupskip,acronymlists={hidden}]{glossaries}
\newlength\fheight
\newlength\fwidth
\def\BibTeX{{\rm B\kern-.05em{\sc i\kern-.025em b}\kern-.08em
    T\kern-.1667em\lower.7ex\hbox{E}\kern-.125emX}}
\begin{document}

\input{acronyms}



\title{Space-O-RAN: Enabling Intelligent, Open, and Interoperable Non Terrestrial Networks in 6G


}

\author{\IEEEauthorblockN{
Eduardo Baena, 
Paolo Testolina, 
Michele Polese, 
Dimitrios Koutsonikolas, 
Josep Jornet, 
Tommaso Melodia
}

\IEEEauthorblockN{Institute for the Wireless Internet of Things, Northeastern University, Boston, MA, U.S.A.}
}

\maketitle

\glsresetall
\glsunset{rapp}
\glsunset{xapp}
\glsunset{dapp}

\begin{abstract}
Satellite networks are rapidly evolving, yet most \glspl{ntn} remain isolated from terrestrial orchestration frameworks. Their control architectures are typically monolithic and static, limiting their adaptability to dynamic traffic, topology changes, and mission requirements. These constraints lead to inefficient spectrum use and underutilized network capacity. Although \gls{ai} promises automation, its deployment in orbit is limited by computing, energy, and connectivity limitations.

This paper introduces Space-O-RAN, a distributed control architecture that extends Open RAN principles into satellite constellations through hierarchical, closed-loop control. Lightweight \glspl{dapp} operate onboard satellites, enabling real-time functions like scheduling and beam steering without relying on persistent ground access. Cluster-level coordination is managed via \glspl{spaceric}, which leverage low-latency \glspl{isl} for autonomous decisions in orbit.
Strategic tasks, including AI training and policy updates, are transferred to terrestrial platforms \glspl{smo} using digital twins and feeder links.

A key enabler is the dynamic mapping of the O-RAN interfaces to satellite links, supporting adaptive signaling under varying conditions. Simulations using the Starlink topology validate the latency bounds that inform this architectural split, demonstrating both feasibility and scalability for autonomous satellite RAN operations.

\end{abstract}


\begin{picture}(0,0)(0,-450)
  \put(20,0){
    \setlength{\fboxsep}{5pt} 
    \fbox{
      \begin{minipage}{0.8\textwidth}
        \footnotesize
        \centering
        This work has been submitted to the IEEE for possible publication.\\
        Copyright may be transferred without notice, after which this version may no longer be accessible.
      \end{minipage}
    }
  }
\end{picture}

\glsresetall
\glsunset{rapp}
\glsunset{xapp}
\glsunset{dapp}

 \vspace{-1 cm}
\section{Introduction}

The rapid growth of \glspl{ntn}, driven by large \gls{leo} constellations such as Starlink or Project Kuiper, reflects a growing need for global broadband access and connectivity in underserved regions~\cite{abdelsadek2023future, cheng2022service}. While these systems offer wide-area coverage, they are currently operated as closed, vertically integrated architectures, with limited programmability and no support for external orchestration. Despite leveraging standardized spectrum and access technologies, \glspl{ntn} remain isolated from the control plane of terrestrial networks and from each other, lacking the operational flexibility needed for modern service delivery.

More critically, this rigid design prevents fine-grained coordination of radio, computational, and energy resources, resulting in unequal satellite utilization, underuse of \glspl{isl} routing, and inefficient use of feeder links for data egress. In current deployments, control decisions are either centralized on the ground or statically encoded, which becomes increasingly unsustainable as constellations scale. For example, feeder link bottlenecks and constrained on-board compute resources limit both user-plane performance and control-plane responsiveness, especially for applications requiring real-time reaction to orbital topology changes. Furthermore, the growing demand for \gls{ai}-driven services on board either for network management predictive tasks or application requiring satellite sensors competes for the same limited communication and power budgets as user services. Without integrated lifecycle management that includes both training and inference workflows spanning space and ground, \gls{ai} adoption remains limited in practice~\cite{ren2023review, iqbal2023ai}.

Standardization efforts such as \gls{3gpp} ~\cite{cheng2022service} and virtualized architectures~\cite{jia2020virtual} still assume that satellites behave like terrestrial base stations with extended delay. This ignores the challenges of orbital dynamics: shifting topologies, frequent \gls{los} loss, and short-lived ISL connections. To overcome these limitations, Space-O-RAN redefines satellites within the radio access network as intelligent autonomous nodes that actively participate in network control. By executing control loops in orbit, it avoids the delays inherent to ground-satellite communication and facilitates real-time decision making directly onboard. The architecture introduces four key innovations:
\begin{itemize}
  \item SpaceRIC: A constellation-aware control plane that hosts satellite-resident application logic, allowing autonomous local decisions within each cluster. A leader–follower synchronization mechanism ensures consistent behavior in dynamic orbital topologies.
    
    \item Hierarchical closed control loops: Control is distributed across three levels, local (onboard satellite), regional (intra-cluster), and global (terrestrial orchestration), allowing coordinated adaptation under variable latency and intermittent connectivity~\cite{iqbal2023ai}.
    
    \item Flexible disaggregated onboard RAN functions: Satellite-deployed \gls{du}, \gls{cu}, and \glspl{dapp} enable configurable placement and execution of RAN functions based on resource availability and control timescales without centralized infrastructure.

    \item Dynamic gateway and routing flexibility: Control and user-plane traffic are decoupled from static feeder link assignments, allowing real-time gateway selection across multi-hop \glspl{isl}.
\end{itemize}

The article is organized as follows. Section~\ref{sec:background} reviews prior work and highlights the current limitations of 3GPP-based NTN integration. Section~\ref{sec:challenges} outlines the core architectural challenges and presents possible use cases. Section~\ref{sec:architecture} details the proposed architecture. Finally, \cref{sec:future,sec:conclusions} discusses future work and summarizes key technical contributions.

\section{Background and Related Work}
\label{sec:background}

Recent \gls{3gpp} efforts have made substantial progress in defining \gls{ntn} support within standardized systems. Releases~17, 18 and 19 have introduced critical extensions to 5G \gls{nr} for satellite-based access, enabling transparent payload operation, \gls{gnss}-based positioning, and protocol-level harmonization across terrestrial and satellite segments~\cite{3gpp_ntn}. However, several technical limitations remain in current specifications. First, constellation-wide autonomous management is only marginally addressed through initial \gls{son} adaptations and resource model extensions~\cite{3GPP_TS28808}, without mechanisms for distributed scheduling, in-orbit orchestration, or \gls{ai}-driven control under dynamic conditions~\cite{mahboob2023revolutionizing}. Moreover, current specifications restrict autonomous \gls{ran} control during intermittent or degraded feeder link conditions to narrow use cases such as store-and-forward for \gls{iot}, lacking general-purpose frameworks for operational continuity without persistent ground access and failing to address the orchestration complexities introduced by dynamic multi-hop \glspl{isl}. Similarly, ITU Recommendation Y.3207 (2024) introduces the INCA framework for Fixed, Mobile, and Satellite Convergence (FMSC), proposing a centralized control model that spans heterogeneous domains. However, its practical adoption in \gls{leo} constellations remains limited due to assumptions of persistent control plane connectivity, high latency tolerance, and the absence of compute-aware resource management.

In parallel, the use of \gls{ai} for optimization of the satellite network is gaining traction, but faces persistent challenges. Existing solutions either rely on periodic ground-side retraining or assume continuous feedback loops, both of which are infeasible under real-world orbital conditions. Latency fluctuations, data sparsity, and limited onboard compute power obstruct the lifecycle management of \gls{ai} models, particularly in use cases requiring fine-grained inference (e.g., beam tracking or link prediction).

Work on modular and disaggregated \gls{ran} architectures using near-RT and non-RT \glspl{ric} shows promise for dynamic control~\cite{oranntn2025}, yet often assumes static backhaul or ground-centered control loops. For \glspl{ntn}, this creates a mismatch between architectural assumptions and operational realities: feeder links are bottlenecks, intra-constellation routing is volatile, and coordination must operate with degraded or disconnected ground access. Fully centralized control becomes impractical on scale and \gls{ran} decisions such as handovers, routing updates, or beam reconfiguration must be handled in situ with limited context. Although standardization bodies and recent proposals provide essential building blocks for \gls{ntn}-\gls{tn} control plane integration, they fall short of enabling decentralized, intelligent, and adaptive control in dynamic satellite constellations.

 \begin{figure*}[h]
    \centering
    \includegraphics[width=0.77\textwidth]{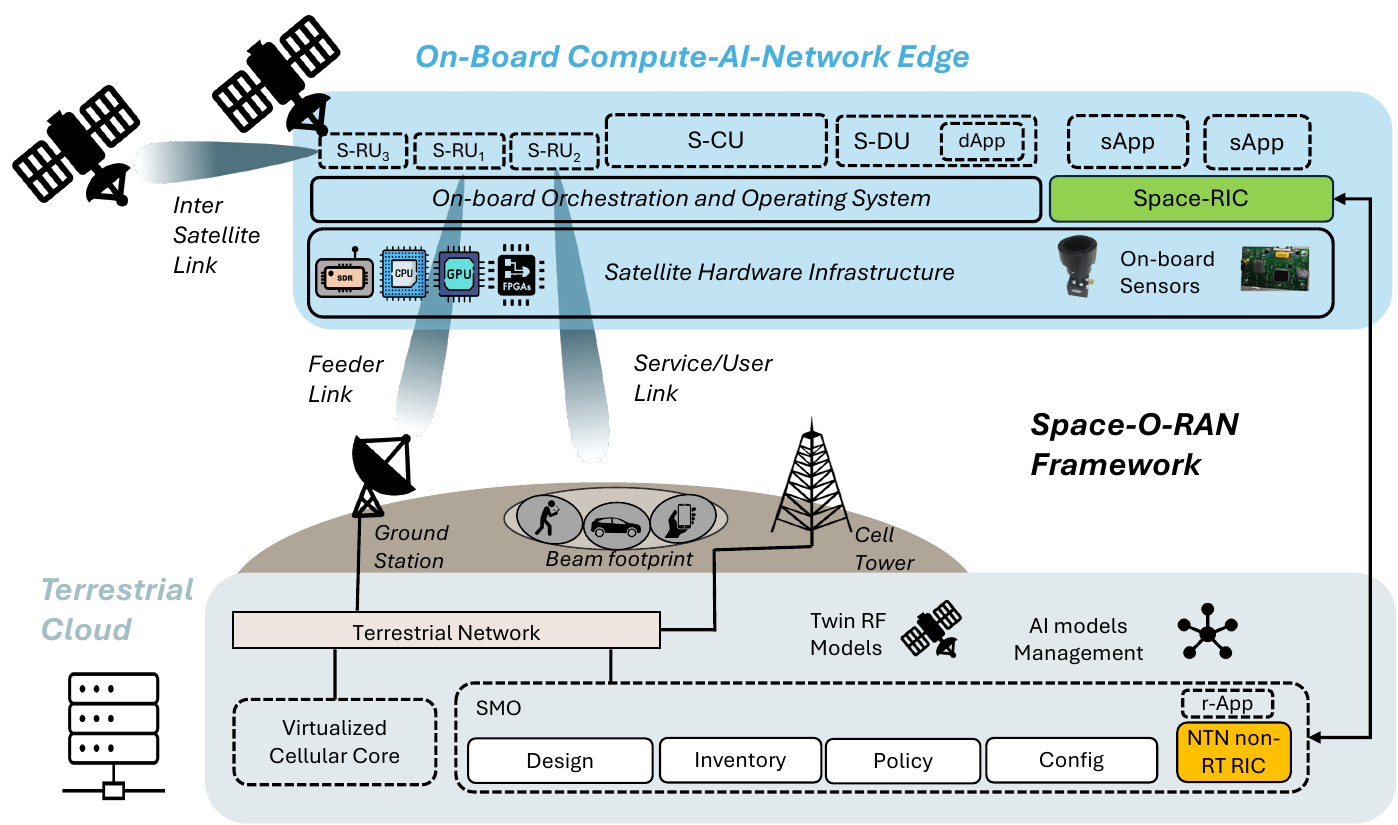}
    \captionsetup{singlelinecheck=false}
    \caption{Proposed Space-O-RAN architecture with partitioned control and shared infrastructure satellite and terrestrial operators}
    \label{fig:oran_space_arch}
\end{figure*}
\vskip -2\baselineskip plus -1fil

\section{Challenges and Use Cases}

Integrating \glspl{ntn} with terrestrial infrastructure involves managing a distributed, dynamic, and resource-constrained communication-computation environment. Traditional centralized architectures and standardized protocols\cite{abdelsadek2023future,cheng2022service} focus primarily on direct-to-device connectivity and regenerative payload support without addressing autonomous coordination and resource-aware operations across satellite constellations. This section outlines major challenges and architectural design requirements.

\subsection{Challenges}
\label{sec:challenges}

\textbf{Decentralized Control in Dynamic Topologies.} The high mobility of \gls{leo} satellites requires frequent handovers and continuous network reconfiguration, causing dynamic latency, intermittent connectivity, and link failures~\cite{ren2023review}. Such conditions preclude conventional centralized management and mandate decentralized federated control. Decisions on handover, beam steering, and routing must occur autonomously and adaptively onboard satellites.

\textbf{Unified Control Abstractions and Standardization Gaps.} Current standards (\gls{3gpp}, ITU-T Y.3207~\cite{cheng2022service}) primarily address user-plane compatibility without providing robust mechanisms for autonomous onboard management or adaptive multi-layer orchestration. There is a lack of standardized control abstractions capable of handling constellation-wide resource allocation and cross-layer decisions under operational dynamics.

\textbf{Constrained AI Lifecycle Management.} Deploying \gls{ai} in \glspl{ntn} faces substantial challenges due to limited computing resources on board, power constraints and intermittent connectivity~\cite{iqbal2023ai}. Effective integration requires a lifecycle-sensitive orchestration that balances data collection, model inference on-board, and model training updates in terrestrial infrastructures without interfering with mission-critical user traffic.

\textbf{Communication-Compute Resource Coupling.} In satellite constellations, communication resources, computational capacity, and energy are closely interdependent and constrained~\cite{wang2023satellite}. Control signaling, telemetry data, and user-plane traffic share the same limited satellite links, necessitating intelligent scheduling and resource allocation policies that jointly optimize these overlapping demands.

\textbf{Flexible Ground Egress and Infrastructure Reuse.} Dependence on fixed ground gateways limits current satellite networks, creating congestion and under-utilization issues~\cite{jia2020virtual}. Efficient use of infrastructure requires flexible and dynamic routing across available \glspl{isl} and gateways, optimizing resource usage by balancing traffic loads and ensuring robust network connectivity even when some feeder links are congested or compromised.

\textbf{Security and Resilient Operation.} Integrating terrestrial and non-terrestrial segments enlarges the attack surface, introducing vulnerabilities in multi-hop communications, authentication processes, and control plane security~\cite{mahboob2023revolutionizing}. Traditional security protocols are insufficient under frequent link interruptions and highly dynamic routing conditions. Ensuring secure operation requires novel federated authentication mechanisms, secure handover strategies, and robust encryption schemes adapted specifically for space-terrestrial interactions.

\textbf{Application-Aware Infrastructure.} Current satellite architectures cannot dynamically instantiate or adapt application-specific network slices or workloads. The ability of satellites to function as intelligent edge platforms requires integrated management of communication, computational and application resources~\cite{chen2025spacegroundfluidai6g}. This integration supports prioritized traffic handling, mission-specific processing, and dynamic service deployment in orbit, transforming satellites into multifunctional and adaptive infrastructure nodes.

Addressing these challenges requires a novel architectural approach that combines decentralized control, federated resource orchestration, secure adaptive communications, and \gls{ai} enabled predictive semi-autonomous operation \cite{DTNTN}.

\begin{figure*}[h]
    \begin{subfigure}[t]{.9\textwidth}
        \centering
        \setlength{\fheight}{0.1\columnwidth}
        \setlength{\fwidth}{\columnwidth}
        \input{figures/ISL_duration.tikz}
       \captionsetup{justification=centering}
        \caption{\gls{pdf} of the \gls{isl} duration in minutes. We defined an \gls{isl} as a link between two satellites with a \gls{rtt} lower than 10~ms, and of duration greater than one minute. Most of the \gls{isl} have a duration of about 5 minutes, but the zoomed-in plot shows the existence of several long-lived (above one hour) \glspl{isl}.}
        \label{fig:isl_pdf}
    \end{subfigure}\\
    \begin{subfigure}[t]{\columnwidth}
        \centering
        \setlength{\fheight}{0.4\columnwidth}
        \setlength{\fwidth}{0.8\columnwidth}
        \input{figures/num_sat}
        \captionsetup{justification=centering}
        \caption{Average number of available \glspl{isl} available at each satellite (solid line), that experience a latency below 10~ms (blue) and 100~ms (orange), in the Starlink constellation.
        The shaded area represents the corresponding standard deviation.}
        \label{fig:isl_latency_num_sat}
    \end{subfigure}
    \begin{subfigure}[t]{\columnwidth}
        \centering
        \setlength{\fheight}{0.4\columnwidth}
        \setlength{\fwidth}{0.8\columnwidth}
       \input{figures/latency_g2s.tikz}
       \captionsetup{justification=centering}
        \caption{\gls{ecdf} of the \gls{rtt} between 33 ground stations and \gls{leo} satellites. The current (2024) Starlink constellation (yellow) orbits at around 550~km.
        Keeping the same constellation configuration, we varied its altitude between the minimum (180~km, blue) and maximum (2000~km, orange) \gls{leo} altitude.}
        \label{fig:g2s_latency}
    \end{subfigure}
    \caption{Duration (a) and number (b) of inter-satellite links below the latency thresholds, and propagation \gls{rtt} for ground-to-satellite (c), as simulated using the November 2024 Starlink constellation (6545 satellites) over 12 hours (12 AM-12 PM, Nov. 11th, 2024, 60 s sampling time). }
    \label{fig:simulation_results}
     \vspace{-10pt}
\end{figure*}

 \vspace{-10pt}
\subsection{Use Cases}

Addressing the abovementioned challenges opens the door to new classes of applications that cannot be effectively supported by current architectures.

\textbf{On-Demand Remote Network Provisioning.} In sparsely populated or infrastructure-poor regions, existing \gls{ntn} systems are unable to flexibly instantiate localized connectivity due to rigid control loops, static scheduling, and dependence on fixed feeder links. Overcoming these constraints would enable on-demand deployment of private network slices supported by orbital assets. This includes temporary 5G coverage in maritime or rural zones without relying on continuous backhaul or pre-positioned terrestrial support~\cite{cheng2022service}.
Such capabilities could significantly reduce deployment costs and extend service reach while minimizing energy consumption and idle resource utilization.

\textbf{Autonomous and Resilient Disaster Response.} Disasters frequently sever terrestrial links, rendering existing cellular systems non-operational. While satellites offer global coverage, current frameworks lack the real-time adaptability to dynamically re-route critical traffic through unaffected regions. Solving the challenges of multi-hop routing, in-situ scheduling, and secure, autonomous link reconfiguration would enable real-time emergency coordination, high-priority backhaul for edge devices, and telemetry aggregation for search-and-rescue operations, even under partial system degradation~\cite{ren2023review}.

\textbf{Distributed AI Workflows and Space-Edge Intelligence.} Global-scale \gls{ai} applications, such as environmental monitoring or predictive control, require distributed data ingestion, adaptive model retraining, and low-latency inference. Traditional architectures are ill-suited for such workflows, as they rely on periodic cloud-side updates and assume deterministic connectivity. Enabling joint orchestration of compute and communication resources, both in the satellite and terrestrial segments, would support federated training and inference pipelines capable of adapting to dynamic link conditions and resource availability~\cite{mahboob2023revolutionizing, oranntn2025}. This is essential to enable true satellite time-sensitive operation autonomy and AI-based applications such as wildfire detection or autonomous fleet management.

\textbf{Secure Multi-Domain Coordination.} Integrating \glspl{ntn} into terrestrial control planes introduces novel security risks, particularly when routing control plane traffic over variable-delay multi-hop satellite paths. Existing protocols assume stable authentication, time synchronization, and trusted infrastructure, which are difficult to guarantee in orbital contexts. Addressing these gaps would enable secure role assignment (e.g., leadership among satellites), robust policy enforcement under intermittent connectivity, and protected telemetry exchange across orbital and terrestrial links~\cite{mahboob2023revolutionizing}.

\vspace{-0.4 cm}
\section{Space-O-RAN Architecture}
\label{sec:architecture}

The Space-O-RAN architecture implements a latency-aware, distributed control system tailored to the operational constraints of dynamic satellite constellations. Designed to operate in environments where persistent ground connectivity is neither available nor desirable, its control logic is hierarchically structured into three layers: onboard execution, intra-cluster coordination, and asynchronous terrestrial orchestration. Each layer corresponds to specific latency bounds, computational capabilities, and resilience requirements. This layered design is illustrated conceptually by the system architecture (\cref{fig:oran_space_arch}), where onboard functions are handled by the Orchestration and Operating System, cluster coordination occurs via \glspl{isl}, and longer-term orchestration is delegated to the \gls{smo} and non-RT \gls{ric}.

At the physical level, each satellite node integrates modular \gls{ran} components designed for autonomous in-orbit operation. The \gls{sru} manages radio connectivity over multiple link types, including \glspl{sl}, \glspl{fl}, \glspl{gsl}, and \glspl{isl}. The \gls{sdu}, a lightweight DU variant adapted for constrained onboard compute, executes core MAC tasks such as scheduling, HARQ, and buffer management. The \gls{scu} oversees inter-node signaling and real-time control execution. Together, these components support \gls{ai}-assisted onboard deployment of \glspl{dapp} responsible for time-critical tasks like beam switching, power control, or link monitoring. These operations, constrained by sub-10~ms timescales, require low-latency execution and are tightly coupled with the radio chain. To support additional services, each node may instantiate virtualized stacks (e.g., NB-IoT, RedCap) and optionally interface with a shared or tenant-specific onboard 5G core.

To enable coordination beyond single-node autonomy, the constellation is logically segmented into control clusters formed based on \gls{isl} connectivity and orbital proximity \cite{clustering}. The viability of this design was evaluated through a full-constellation simulation of Starlink (6545 \glspl{leo}) using the MATLAB Satellite Communications Toolbox. Satellite trajectories were propagated over a 12-hour window (Nov. 11, 2024), at 60-second resolution, using publicly available \glspl{tle}. As shown in \cref{fig:isl_latency_num_sat}, each satellite maintains low-latency ISLs with over 420 (375) peers on average at below 10~ms (blue) and 100~ms (orange) one-way delay, respectively. Despite rapid topology shifts, the constellation's density and shell organization enable sufficient link stability for coordinated operation.

To assess ISL persistence, we analyzed the empirical \gls{pdf} of link durations for those sustaining round-trip time (RTT) under 10~ms, discarding intervals shorter than one minute. Results in \cref{fig:isl_pdf} reveal most viable links last about 5 minutes, with a long tail extending to multiple hours. These findings confirm the feasibility of stable control clusters capable of sustaining reactive coordination over bounded-latency links.

Within each cluster, a distributed control plane is instantiated through the \gls{spaceric}, which acts as the in-orbit analog to a non-RT \gls{ric}. The \gls{spaceric} runs energy efficient \glspl{sapp}, adapted from terrestrial \glspl{xapp}, to manage local coordination tasks, for example, beam handovers, spectrum sharing, resource alignment or policy enforcement. Each \gls{sapp} operates on shared state and propagates decisions via intra-cluster signaling to preserve alignment even under node mobility or partial link failures.

These interface mappings enable multiple control loops to coexist over shared physical links, each with distinct latency and reliability requirements. Near real-time signaling (e.g., over \gls{e2}, between near-RT \gls{ric} and RAN nodes) is prioritized over short-delay paths via \glspl{isl}, allowing rapid adaptation to beam misalignment or interference. Less time-sensitive exchanges (e.g., model updates via \gls{a1}, from non-RT to near-RT \gls{ric}) tolerate delay and are scheduled opportunistically when bandwidth and power permit. F1 (CU–DU), \gls{o1} (RAN configuration and telemetry), and O2 (resource orchestration and lifecycle management) are routed based on timing and reliability demands. Interface coexistence is managed through logical separation and onboard schedulers that prioritize critical control during congestion or eclipse. The detailed mapping between O-RAN interfaces and physical link types is summarized in \cref{tab:latency_alignment_space_oran}.

Long-term orchestration is offloaded to the terrestrial \gls{smo} and cloud-based \glspl{ai} pipelines, which maintain a global \gls{dt} of the constellation and issue periodic strategic updates through \glspl{gsl}. Given the variability in \gls{gsl} latency, from 1.2~ms to over 37~ms depending on altitude and geometry, these updates are used for policy synchronization and model refinement, not real-time decisions. This separation ensures that satellites remain responsive and self-sufficient during extended ground disconnects.

\begin{table*}[h]
\centering
\scriptsize
\setlength{\tabcolsep}{2pt}
\caption{Dynamic Interface-Link Mapping}
\label{tab:latency_alignment_space_oran}
\begin{tabular}{|p{2.5cm}|p{1.8cm}|p{4cm}|p{4cm}|p{2.5cm}|p{2.2cm}|}
\hline
\textbf{O-RAN Interface} & \textbf{Link Type} & \textbf{Primary Function} & \textbf{Notes} & \textbf{Typical Latency Target}  & \multirowcell{5}{\textbf{RTT (Starlink)}\\ (P5/P50/P95) \\ GSL: 5/12.3/16.7 ms \\ ISL: 4/14.7/21.7 ms}\\
\cline{1-5}
\textbf{Midhaul (F1-C/U)} & ISL / FL / GSL & DU–CU communication & CU functions over FL/GSL may impair HARQ timing, dynamic scheduling, or beam switching responsiveness. & 1.5–10 ms  & \\
\cline{1-5}
\textbf{Backhaul / Core (NG, N2/N3)} & SL / FL / GSL & CU–5GC or core-plane communication & The links can tolerate moderate delay in NTN context & $<$50–100 ms \newline (service-dependent) & \\
\cline{1-5}
\textbf{Near-RT Control (E2)} & ISL / FL / SL & Policy, RIC feedback loops & Fast loops feasible over ISL only; SL/FL need fallback to slower control &  $<$10–20 ms &\\
\cline{1-5}
\textbf{Management (O1, A1)} & FL / GSL & SMO, policy, AI model sync & Not delay-sensitive; supports asynchronous operation & Tolerant ($>$100 ms) & \\
\hline
\end{tabular}
\end{table*}

\begin{figure}[h]
\centering
\includegraphics[width=0.9\linewidth, trim=0 0 100 0, clip]{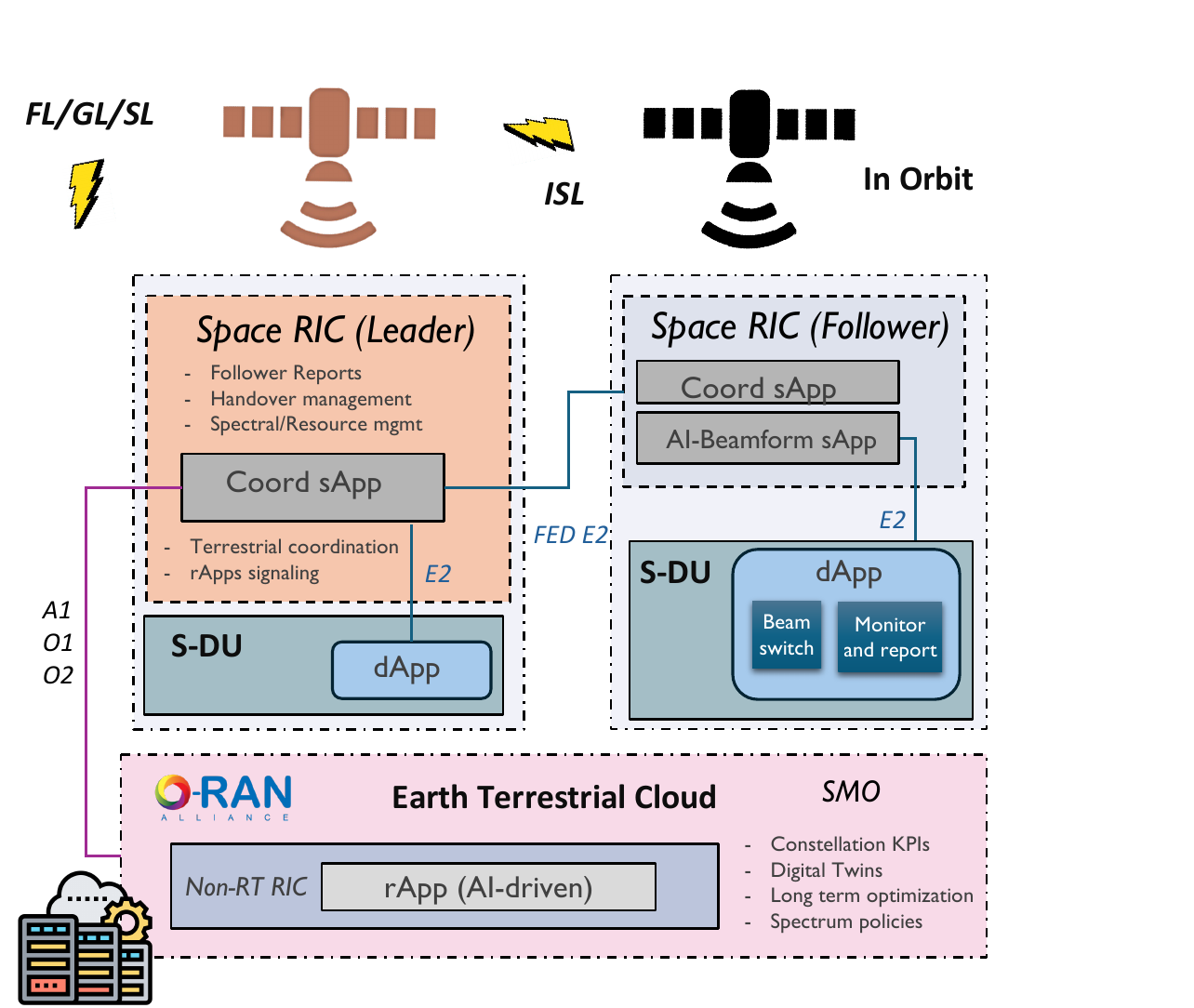}
\captionsetup{justification=centering}
\caption{Hierarchical coordination workflow}
\label{fig:spaceRIC}
\vspace{-0.3 cm}
\end{figure}

To manage the tight coupling between communication, compute, and energy resources, the onboard operating system dynamically instantiates and scales virtualized RAN functions, \glspl{sapp}, and \glspl{dapp} based on current link conditions, traffic load, and energy availability. This mechanism also supports application-aware execution~\cite{chen2025spacegroundfluidai6g}, enabling mission-specific services to run as \glspl{sapp} (i.e. Earth Observation) directly integrated into the transmission loop, without relying on external orchestration.

\vspace{-0.3 cm}
\subsection{Onboard Autonomy and Cluster Coordination}

Each coordination cluster operates as a self-contained control domain instantiated via a distributed \gls{spaceric}, where nodes collectively execute the functions required for intra-cluster operation. The \gls{spaceric} hosts multiple \glspl{sapp} tailored to orbital execution constraints, similar to terrestrial \glspl{xapp}, managing tasks such as spectrum reuse to avoid interferences with other constellations and \gls{ai}-based resource optimization.

At the core of the cluster, a dedicated Coordination \gls{sapp} (Coord sApp), shown in Figure~\ref{fig:spaceRIC}, synchronizes the behavior of follower nodes, orchestrates the operation of other \glspl{sapp}, and interfaces with the terrestrial control plane via asynchronous policies and model updates. It acts as the control anchor of the cluster, ensuring a coherent execution across drifting nodes and variable links.

The \gls{spaceric} employs a Leader–Follower architecture, where one satellite temporarily assumes coordination duties, aggregating local metrics, managing inter-satellite handovers, and disseminating control actions. The followers maintain soft-state replicas of control parameters and execute delegated logic to ensure continuity. Leadership transitions are governed by \gls{fede2}, an enhanced failover protocol built over the standard \gls{e2} interface. Beyond encoding leader election metadata, \gls{fede2} provides tolerance to link disruptions by allowing nodes to securely rejoin the cluster and resume synchronization once \gls{isl} connectivity is restored. This design enables stateful reconfiguration and continuity even in the presence of intermittent \gls{isl}  dropouts, without requiring ground participation or manual recovery. 

Cluster-wide state is continuously synchronized over low-latency \glspl{isl}, which maintain bounded message delays in the 5–10~ms range even under orbital drift. In Figure~\ref{fig:spaceRIC}, the left satellite hosts the Coord \gls{sapp} and acts as cluster leader, receiving \gls{ai} model updates from the terrestrial \gls{rapp} via \gls{a1}/\gls{o1}. The right node operates as a follower, executing \gls{ai} driven \glspl{sapp} (e.g., beamforming, monitoring), whose decisions are enacted through \glspl{dapp} within \gls{sdu}.

The closed-loop sequence example (\cref{fig:spaceRIC} begins with CSI-driven \gls{dapp} decisions at the RAN layer, propagates through the \gls{spaceric} for cluster-wide coherence, and if persistent anomalies are detected, triggers feedback to the \gls{smo} for non-RT policy adjustment. In this regard, continuous autonomy at the cluster level, even during extended disconnection from the terrestrial \gls{smo} is supported. Upon \gls{gsl} restoration, updated models and strategic policies are propagated to the cluster leader and distributed locally without interrupting real-time responsiveness. Clusters are dynamically formed during orbital grouping events or in mission-defined coordination phases. A leader is elected based on a weighted score that combines \gls{isl} connectivity, compute availability, and telemetry freshness. \gls{fede2} ensures continuity by enabling logical control to migrate to synchronized followers without external input. To prevent partitioned decision-making, a minimum quorum of state-synchronized nodes is required to sustain cluster-level coordination; otherwise, nodes revert to local-only operation.

\vspace{-0.5 cm}
\subsection{Ground-Based Orchestration and AI Pipelines}
\glsreset{dt}

The terrestrial segment, jointly operated with the satellite provider, is responsible for global orchestration, policy generation, and \gls{ai} lifecycle management across satellite clusters. These functions are distributed between two main components: the \gls{smo}, which collects telemetry, configures infrastructure, and monitors system health via the O1/O2 interfaces; and the non-RT \gls{ric}, which hosts \glspl{rapp} that drive long-term optimization using cloud-scale \gls{ai} pipelines. \glspl{rapp} implement domain-specific analytics and control logic, transforming aggregated data into adaptive policies. These may include time-evolving spectrum coordination, congestion mitigation strategies, mobility forecasts, or slice prioritization rules. Each \gls{rapp} operates asynchronously, combining historical traces and real-time inputs to refine predictive models and generate control strategies over medium- to long-term horizons.

These \gls{ai} pipelines also serve as the backend for training two distinct but interlinked systems: digital twins (\glspl{dt}) that emulate constellation-wide network behavior, and deployable inference models executed onboard each satellite. The \glspl{dt} integrate orbital geometry, historical \glspl{kpi}, and link performance data to simulate coordination bottlenecks, capacity fluctuations, and potential failure modes, providing a predictive foundation for global orchestration. Meanwhile, on-board models derived from these same pipelines enable localized real-time decision making with minimal ground interaction.

High-level policies and updated models are transmitted to \gls{spaceric} via the \gls{a1} interface. These policies encode intent-based directives, for example spectrum reuse plans or  energy budget swhich are interpreted by in-orbit \glspl{sapp} and enforced by co-located \glspl{dapp} at the RAN level.

Satellite telemetry used to feed these pipelines is compressed and selectively sampled to align with \gls{gsl} bandwidth constraints, enabling asynchronous yet coherent model updates. To prevent congestion on feeder and gateway links, model updates are prioritized using anomaly scores computed locally at the cluster level. Only clusters exhibiting significant behavioral divergence initiate full-state uploads or receive complete model retraining. The rest continue operating with cached models, periodically refined through lightweight differential updates.

\vspace{-0.5 cm}
\subsection{Security and Coordination Integrity}

Secure coordination is maintained through cryptographically grounded protocols that minimize the attack surface while ensuring responsiveness. Each node authenticates its control role via mutual certificate-based validation, with critical leadership functions assigned through a quorum-based election enforced by the \gls{fede2} protocol. This federated extension of the \gls{e2} interface restricts leadership promotion to candidates with verified cryptographic credentials and synchronized telemetry history.

All control interfaces (\glspl{o1}, \glspl{a1}, \glspl{e2}) operate over lightweight encrypted channels using ephemeral session keys, derived from pre-distributed roots and rotated periodically to prevent replay and downgrade attacks. The onboard \gls{scu} integrates a priority-aware scheduler that ensures authentication flows are prioritized, avoiding delays during high traffic or contention. Rapid link context switching is supported through pre-authentication of adjacent nodes, allowing seamless handovers without compromising integrity.

In adversarial conditions, for example, contested spectrum or suspected node compromise, satellites automatically reduce their control-plane exposure by disabling downstream reconfiguration via \gls{a1} and rejecting unsigned or tampered updates at the \gls{sapp} layer. Leadership failover incorporates real-time integrity checks: if the acting Leader becomes unresponsive or diverges from expected behavior, a cluster-wide consensus process is triggered, promoting a valid Follower using cached keys and buffered telemetry to ensure continuity within a single control cycle. 

\vspace{-0.5cm}
\section{Future Outlook}
\label{sec:future}

Extending the proposed framework to lunar and deep-space scenarios introduces new architectural imperatives. The intermittency and propagation delays intrinsic to such environments challenge traditional coordination schemes and demand resilient, locally autonomous control structures. While protocols like the Bundle Protocol and \gls{hdtn}~\cite{dudukovich2024advances} offer improved delay-tolerant communication, they are not designed to support the real-time orchestration loops required for joint spectrum, compute, and queue arbitration under dynamic mission constraints.

Future lunar networks may integrate \glspl{spaceric} directly within surface assets to support close-proximity swarm coordination and shared control between landers, orbiters, and surface stations. This would enable fast reactive decisions within clusters while preserving synchronization with Earth-based or Lunar Base orchestration. In such settings, the use of steerable \glspl{isl} and directional spectrum reallocation must be extended to heterogeneous assets, which can be operated at FR3/THz frequencies to achieve low-latency high-bandwidth communication in short-range vacuum links\cite{Thz2}. However, these links require dynamic adaptation to occlusions and relative motion, reinforcing the need for on-board arbitration mechanisms that prioritize control stability over throughput.

In addition, space deployments will benefit from the introduction of a semantic control network layer capable of interpreting mission context, task intent, and data relevance to reduce protocol overhead and improve autonomy. The use of AI accelerators and specialized inference processors, designed for space-grade reliability and low-power operation, becomes critical to sustain in situ adaptation, with federated learning offering a path for collaborative inference without raw data exchange.

\vspace{-0.3cm}
\section{Conclusions}
\label{sec:conclusions}

This work presented Space-O-RAN, a distributed control architecture that addresses a fundamental limitation in current \gls{ntn} deployments: the inability to execute timely, adaptive decisions due to latency, intermittent connectivity, and on-board resource constraints. A three-tier control hierarchy aligned with physical link characteristics and onboard processing capabilities. Real-time decisions are handled autonomously in orbit by lightweight AI-driven apps, co-located with the radio stack, and tailored to local context. Cluster-level coordination is managed by the \gls{spaceric}, enabling federated alignment and resilient operation even under partial link degradation. Long-term strategy and model training remain on the Terrestrial cloud, but inference is executed onboard, allowing for continuous adaptation without waiting for terrestrial input.

A key enabler is the dynamic mapping of O-RAN interfaces to available satellite links, allowing control loops to persist transparently over shared \gls{isl}, \gls{gsl}, or feeder paths depending on latency and policy requirements. Simulation results validate the feasibility of maintaining sub-10~ms intra-cluster coordination over real-world ISL dynamics, while asynchronous updates from terrestrial digital twins enable refined policy evolution and \gls{ai} model improvement over time. Together, these mechanisms offer a scalable foundation for autonomous, intelligent, and resilient \glspl{ntn} beyond traditional architectures.

\vspace{-0.4cm}
\bibliographystyle{IEEEtran}
\bibliography{main}

\vskip -3\baselineskip plus -1fil

\begin{IEEEbiographynophoto}{Eduardo Baena} is a postdoctoral research fellow at Northeastern University, with a Ph.D. in Telecommunication Engineering from the University of Malaga. His experience includes roles in the international private sector (2010–2017) and as Co-PI in several research projects.
\end{IEEEbiographynophoto}

\vskip -2\baselineskip plus -1fil

\begin{IEEEbiographynophoto}{Paolo Testolina} received a Ph.D. in information engineering from the University of Padova in 2022. He is a Research Scientist at Northeastern University, with research interests in mmWave and sub-THz networks, including channel modeling, physical layer simulation, Non-Terrestrial Networks and spectrum sharing and coexistence.
\end{IEEEbiographynophoto}

\vskip -2\baselineskip plus -1fil

\begin{IEEEbiographynophoto}{Michele Polese} is a Research Assistant Professor at Northeastern University’s Institute for the Wireless Internet of Things. He received his Ph.D. from the University of Padova in 2020 and focuses on protocols for 5G and beyond, mmWave/THz networks, spectrum sharing, and open RAN development.
\end{IEEEbiographynophoto}

\vskip -2\baselineskip plus -1fil

\begin{IEEEbiographynophoto}{Dimitrios Koutsonikolas} is an Associate Professor in the department of Electrical and Computer Engineering and a member of the Institute for the Wireless Internet of Things at Northeastern University. He specializes in experimental wireless networking and mobile computing, with a focus on 5G/NextG networks and applications. He is a recipient of the IEEE Region 1 Technological Innovation Award and the NSF CAREER Award.
\end{IEEEbiographynophoto}

\vskip -2\baselineskip plus -1fil

\begin{IEEEbiographynophoto}{Josep M. Jornet} (M'13--SM'20--F'24) is a Professor and Associate Director of the Institute for the Wireless Internet of Things at Northeastern University. His research focuses on terahertz communications and networking, pioneering advancements in nanoscale communications and wireless systems. He has authored over 250 peer-reviewed publications and holds five US patents.
\end{IEEEbiographynophoto}

\vskip -2\baselineskip plus -1fil

\begin{IEEEbiographynophoto}{Tommaso Melodia} is the William Lincoln Smith Chair Professor at Northeastern University and Founding Director of the Institute for the Wireless Internet of Things. A recipient of the NSF CAREER award, he has served as an Associate Editor for leading IEEE journals.
\end{IEEEbiographynophoto}

\end{document}

%% file: acronyms.tex
\newacronym{3gpp}{3GPP}{Third Generation Partnership Project}
\newacronym{5g}{5G}{Fifth-Generation}
\newacronym{6g}{6G}{Sixth-Generation}
\newacronym{a1}{A1}{O-RAN Policy Management Interface}
\newacronym{ai}{AI}{Artificial Intelligence}
\newacronym{awgn}{AWGN}{Average White Gaussian Noise}
\newacronym{cpu}{CPU}{Central Processing Unit}
\newacronym{cu}{CU}{Centralized Unit}
\newacronym{daa}{DAA}{Decentralized Authentication Authority}
\newacronym{dapp}{dApp}{Distributed RAN Application}
\newacronym{dt}{DT}{Digital Twin}
\newacronym{du}{DU}{Distributed Unit}
\newacronym{e2}{E2}{O-RAN Near-Real-Time Interface}
\newacronym{ecdf}{ECDF}{Empirical Cumulative Distribution Function}
\newacronym{fede2}{FED-E2}{Federated O-RAN Near-Real-Time Interface}
\newacronym{fl}{FL}{Feeder Link}
\newacronym{geo}{GEO}{Geostationary Orbit}
\newacronym{gnss}{GNSS}{Global Navigation Satellite System}
\newacronym{gpu}{GPU}{Graphics Processing Unit}
\newacronym{gs}{GS}{Geometric Spread}
\newacronym{hap}{HAP}{High Altitude Platform}
\newacronym{hdtn}{HDTN}{High-Delay Tolerant Networking}
\newacronym{iot}{IoT}{Internet of Things}
\newacronym{iq}{IQ}{In-phase and Quadrature}
\newacronym{iab}{IAB}{Integrated Access Backhaul}
\newacronym{isl}{ISL}{Inter-Satellite Link}
\newacronym{leo}{LEO}{Low-Earth Orbit}
\newacronym{los}{LOS}{Line-Of-Sight}
\newacronym{meo}{MEO}{Medium Earth Orbit}
\newacronym{mec}{MEC}{Mobile Edge Computing}
\newacronym{ml}{ML}{Machine Learning}
\newacronym{mmwave}{mmWave}{millimeter wave}
\newacronym{mppt}{MPPT}{ Maximum Power Point Tracking }
\newacronym{near-rt-ric}{near-RT-RIC}{near-RT-\gls{ric}}
\newacronym{nr}{NR}{New Radio}
\newacronym{ntn}{NTN}{Non-Terrestrial Network}
\newacronym{o1}{O1}{O-RAN Operations and Management Interface}
\newacronym{oran}{O-RAN}{O-RAN ALLIANCE}
\newacronym{os}{OS}{Operating System}
\newacronym{osiran}{OSIRAN}{Open-Space-Integrated-RAN}
\newacronym{pe}{PE}{Pointing Error}
\newacronym{pdf}{PDF}{Probability Density Function}
\newacronym{qos}{QoS}{Quality of Service}
\newacronym{ran}{RAN}{Radio Access Network}
\newacronym{ric}{RIC}{RAN Intelligent Controller}
\newacronym{rapp}{rApp}{RAN Application}
\newacronym{rtt}{RTT}{Round-Trip Time}
\newacronym{sapp}{sApp}{Space Application}
\newacronym{sagin}{SAGIN}{Space-Air-Ground Integrated Network}
\newacronym{sl}{SL}{Service Link}
\newacronym{smo}{SMO}{Service Management and Orchestration}
\newacronym{spaceran}{Space-RAN}{Space integrated Open Radio Access network}
\newacronym{spaceric}{Space-RIC}{Space RAN Intelligent Controller}
\newacronym{sru}{s-RU}{Satellite Radio Unit}
\newacronym{sdu}{s-DU}{Satellite Distributed Unit}
\newacronym{son}{SON}{Self Organizing Network}
\newacronym{scu}{s-CU}{Satellite Central Unit}
\newacronym{snf}{s-NF}{Satellite Network Function}
\newacronym{tn}{TN}{Terrestrial Network}
\newacronym{tle}{TLE}{Two-Line Element Set}
\newacronym{ue}{UE}{User Equipment}
\newacronym{ul}{UL}{User Link}
\newacronym{xapp}{xApp}{eXtended RAN Application}
\newacronym{gsl}{GSL}{ground-to-satellite link}
\newacronym{kpi}{KPI}{Key Performance Indicator}

%% file: figures/ISL_duration.tikz
%
%
\definecolor{mycolor1}{rgb}{0.00000,0.44700,0.74100}%
\begin{tikzpicture}

\begin{axis}[%
width=0.951\fwidth,
height=\fheight,
at={(0\fwidth,0\fheight)},
scale only axis,
xmin=0,
xmax=721,
xlabel style={font=\color{white!15!black}},
xlabel={ISL duration [m]},
ymin=0,
ymax=0.25,
ylabel style={font=\color{white!15!black}},
ylabel={PDF},
axis background/.style={fill=white},
xmajorgrids,
ymajorgrids
]
\addplot[ybar interval, fill=mycolor1, fill opacity=0.6, draw=black, area legend] table[row sep=crcr] {%
x	y\\
1	0.0387743905356456\\
2	0.0826699163234119\\
3	0.161899522650716\\
4	0.247137780673188\\
5	0.152066521753234\\
6	0.0737282482320875\\
7	0.0491085011113744\\
8	0.0328568368413602\\
9	0.0233551881371604\\
10	0.0190252414962772\\
11	0.016152862869522\\
12	0.0119992463389593\\
13	0.010601291330064\\
14	0.00978003119099434\\
15	0.00762774374422113\\
16	0.00575238997972875\\
17	0.00516965253498033\\
18	0.00510912077566591\\
19	0.00443139829353444\\
20	0.00473511719097017\\
21	0.00364063372264242\\
22	0.00352720293023186\\
23	0.00324624086466915\\
24	0.00298736423376581\\
25	0.0025262556947735\\
26	0.00237847763438068\\
27	0.00223557603796066\\
28	0.00147470631142357\\
29	0.00131236953250329\\
30	0.00120826762769286\\
31	0.00112706390153727\\
32	0.00123236725399317\\
33	0.00159767801160697\\
34	0.000990098869953681\\
35	0.00068599126886794\\
36	0.000559415225748126\\
37	0.000473865086051567\\
38	0.000403403715314295\\
39	0.000337818808549812\\
40	0.000284637081890035\\
41	0.000215271148711868\\
42	0.000170534892265836\\
43	0.000113536802496919\\
44	7.26169091600277e-05\\
45	3.66441531868363e-05\\
46	1.71736339911306e-05\\
47	6.07791161826022e-06\\
48	5.75988135916521e-06\\
49	4.55843371369517e-06\\
50	5.61853457734521e-06\\
51	5.15915753643019e-06\\
52	5.01781075461018e-06\\
53	4.94713736370018e-06\\
54	5.4771877955252e-06\\
55	5.26516762279519e-06\\
56	5.79521805462021e-06\\
57	5.61853457734521e-06\\
58	5.05314745006519e-06\\
59	5.3711777091602e-06\\
60	6.11324831371522e-06\\
61	4.98247405915518e-06\\
62	5.75988135916521e-06\\
63	4.91180066824518e-06\\
64	5.15915753643019e-06\\
65	5.5478611864352e-06\\
66	5.72454466371021e-06\\
67	6.43127857281024e-06\\
68	5.3358410137052e-06\\
69	4.80579058188018e-06\\
70	5.83055475007521e-06\\
71	6.36060518190023e-06\\
72	6.53728865917524e-06\\
73	9.22287751375534e-06\\
74	1.37106378365405e-05\\
75	2.08839870139058e-05\\
76	1.27565470592555e-05\\
77	2.14847108366408e-05\\
78	2.01065797138957e-05\\
79	2.6325838113976e-05\\
80	4.18386474187215e-05\\
81	5.08848414552019e-05\\
82	3.92590686505064e-05\\
83	2.6255164723066e-05\\
84	2.41349629957659e-05\\
85	2.8446039841276e-05\\
86	2.58664610730609e-05\\
87	2.7809979323086e-05\\
88	2.39582795184909e-05\\
89	2.18380777911908e-05\\
90	1.57248294774756e-05\\
91	1.22971700183405e-05\\
92	8.1274399546503e-06\\
93	6.71397213645025e-06\\
94	6.43127857281024e-06\\
95	4.55843371369517e-06\\
96	4.16973006369015e-06\\
97	4.80579058188018e-06\\
98	5.93656483644022e-06\\
99	4.77045388642518e-06\\
100	5.26516762279519e-06\\
101	5.5125244909802e-06\\
102	4.13439336823515e-06\\
103	5.5478611864352e-06\\
104	4.62910710460517e-06\\
105	6.14858500917023e-06\\
106	6.92599230918025e-06\\
107	7.56205282737028e-06\\
108	7.52671613191528e-06\\
109	9.54090777285035e-06\\
110	1.16611095001504e-05\\
111	1.12724058501454e-05\\
112	9.71759125012536e-06\\
113	1.26858736683455e-05\\
114	1.80217146820507e-05\\
115	1.73149807729506e-05\\
116	2.04246099729908e-05\\
117	2.8163346277636e-05\\
118	3.94357521277814e-05\\
119	5.07434946733819e-05\\
120	2.53010739457809e-05\\
121	4.23686978505466e-05\\
122	3.66441531868363e-05\\
123	1.98592228457107e-05\\
124	1.82690715502357e-05\\
125	2.24388016139258e-05\\
126	1.72796440774956e-05\\
127	1.58661762592956e-05\\
128	1.62548799093006e-05\\
129	2.22974548321058e-05\\
130	2.91174370549211e-05\\
131	3.24744231231462e-05\\
132	3.56547257140963e-05\\
133	3.29338001640612e-05\\
134	2.91174370549211e-05\\
135	2.26508217866558e-05\\
136	1.89758054593357e-05\\
137	1.79157045956857e-05\\
138	1.60075230411156e-05\\
139	1.31805874047155e-05\\
140	1.23325067137955e-05\\
141	1.14844260228754e-05\\
142	1.42760249638205e-05\\
143	8.69282708193032e-06\\
144	8.0567665637403e-06\\
145	8.79883716829532e-06\\
146	9.61158116376035e-06\\
147	1.42760249638205e-05\\
148	9.64691785921536e-06\\
149	7.73873630464528e-06\\
150	8.0214298682853e-06\\
151	8.37479682283531e-06\\
152	1.00356215092204e-05\\
153	1.62195432138456e-05\\
154	2.6290501418521e-05\\
155	2.19440878775558e-05\\
156	3.13083121731311e-05\\
157	4.15559538550815e-05\\
158	2.00005696275307e-05\\
159	1.82690715502357e-05\\
160	3.00715278322061e-05\\
161	3.12023020867661e-05\\
162	9.68225455467036e-06\\
163	9.43489768648535e-06\\
164	2.8587386623096e-05\\
165	2.31455355230258e-05\\
166	1.10250489819604e-05\\
167	1.16611095001504e-05\\
168	1.60428597365706e-05\\
169	1.14490893274204e-05\\
170	1.05303352455904e-05\\
171	1.26858736683455e-05\\
172	1.40993414865455e-05\\
173	1.97885494548007e-05\\
174	3.14143222594962e-05\\
175	3.10962920004011e-05\\
176	3.59020825822813e-05\\
177	4.18033107232665e-05\\
178	3.89763750868664e-05\\
179	3.32164937277012e-05\\
180	3.08135984367611e-05\\
181	2.8587386623096e-05\\
182	2.25094750048358e-05\\
183	1.83750816366007e-05\\
184	1.08483655046854e-05\\
185	6.78464552736025e-06\\
186	5.4418511000702e-06\\
187	4.62910710460517e-06\\
188	4.55843371369517e-06\\
189	4.91180066824518e-06\\
190	3.92237319550514e-06\\
191	3.56900624095513e-06\\
192	3.46299615459013e-06\\
193	3.42765945913513e-06\\
194	3.46299615459013e-06\\
195	3.88703650005014e-06\\
196	3.25097598186012e-06\\
197	3.60434293641013e-06\\
198	3.92237319550514e-06\\
199	3.92237319550514e-06\\
200	3.00361911367511e-06\\
201	3.63967963186513e-06\\
202	3.71035302277514e-06\\
203	3.99304658641515e-06\\
204	3.88703650005014e-06\\
205	4.20506675914515e-06\\
206	3.92237319550514e-06\\
207	4.52309701824017e-06\\
208	4.48776032278517e-06\\
209	4.62910710460517e-06\\
210	5.19449423188519e-06\\
211	5.15915753643019e-06\\
212	6.07791161826022e-06\\
213	8.94018395011533e-06\\
214	1.13430792410554e-05\\
215	2.05306200593558e-05\\
216	1.93291724138857e-05\\
217	2.55837675094209e-05\\
218	2.33222190003009e-05\\
219	3.40999111140763e-05\\
220	2.90821003594661e-05\\
221	1.40286680956355e-05\\
222	1.57955028683856e-05\\
223	1.80923880729607e-05\\
224	1.25445268865255e-05\\
225	1.26505369728905e-05\\
226	1.22618333228854e-05\\
227	1.13784159365104e-05\\
228	1.51241056547406e-05\\
229	2.8269356364001e-05\\
230	2.00005696275307e-05\\
231	1.20144764547004e-05\\
232	8.1627766501053e-06\\
233	6.71397213645025e-06\\
234	4.73511719097017e-06\\
235	3.60434293641013e-06\\
236	4.27574015005516e-06\\
237	3.81636310914014e-06\\
238	4.20506675914515e-06\\
239	3.46299615459013e-06\\
240	3.46299615459013e-06\\
241	3.56900624095513e-06\\
242	3.56900624095513e-06\\
243	3.07429250458511e-06\\
244	3.42765945913513e-06\\
245	4.45242362733016e-06\\
246	5.08848414552019e-06\\
247	6.64329874554024e-06\\
248	5.83055475007521e-06\\
249	6.57262535463024e-06\\
250	9.01085734102533e-06\\
251	1.69969505138556e-05\\
252	1.62902166047556e-05\\
253	1.24385168001605e-05\\
254	1.15197627183304e-05\\
255	1.41700148774555e-05\\
256	1.13077425456004e-05\\
257	9.15220412284534e-06\\
258	1.04596618546804e-05\\
259	1.22618333228854e-05\\
260	2.27921685684758e-05\\
261	3.19090359958662e-05\\
262	9.04619403648033e-06\\
263	3.49833285004513e-06\\
264	5.08848414552019e-06\\
265	6.11324831371522e-06\\
266	5.61853457734521e-06\\
267	1.02123049864954e-05\\
268	2.45590033412259e-05\\
269	6.28993179099023e-06\\
270	6.39594187735523e-06\\
271	7.56205282737028e-06\\
272	8.51614360465531e-06\\
273	7.84474639101029e-06\\
274	9.43489768648535e-06\\
275	8.79883716829532e-06\\
276	8.51614360465531e-06\\
277	7.91541978192029e-06\\
278	6.60796205008524e-06\\
279	4.69978049551517e-06\\
280	4.16973006369015e-06\\
281	2.8622723318551e-06\\
282	2.47356868185009e-06\\
283	2.50890537730509e-06\\
284	3.18030259095012e-06\\
285	2.54424207276009e-06\\
286	1.59015129547506e-06\\
287	2.04952833639008e-06\\
288	1.87284485911507e-06\\
289	1.83750816366007e-06\\
290	1.97885494548007e-06\\
291	2.22621181366508e-06\\
292	2.15553842275508e-06\\
293	2.57957876821509e-06\\
294	2.7915989409451e-06\\
295	3.67501632732013e-06\\
296	3.63967963186513e-06\\
297	3.49833285004513e-06\\
298	3.07429250458511e-06\\
299	4.09905667278015e-06\\
300	4.94713736370018e-06\\
301	4.66444380006017e-06\\
302	2.8269356364001e-06\\
303	2.93294572276511e-06\\
304	3.63967963186513e-06\\
305	2.89760902731011e-06\\
306	2.43823198639509e-06\\
307	2.6149154636701e-06\\
308	3.14496589549512e-06\\
309	3.49833285004513e-06\\
310	3.60434293641013e-06\\
311	3.95770989096015e-06\\
312	4.45242362733016e-06\\
313	4.24040345460016e-06\\
314	3.92237319550514e-06\\
315	4.24040345460016e-06\\
316	6.11324831371522e-06\\
317	6.46661526826524e-06\\
318	6.74930883190525e-06\\
319	7.59738952282528e-06\\
320	6.99666570009026e-06\\
321	6.81998222281525e-06\\
322	8.2334500410153e-06\\
323	7.38536935009527e-06\\
324	9.75292794558036e-06\\
325	7.66806291373528e-06\\
326	4.27574015005516e-06\\
327	3.46299615459013e-06\\
328	2.54424207276009e-06\\
329	1.90818155457007e-06\\
330	1.48414120911005e-06\\
331	2.22621181366508e-06\\
332	2.19087511821008e-06\\
333	1.55481460002006e-06\\
334	1.76683477275006e-06\\
335	2.15553842275508e-06\\
336	1.51947790456506e-06\\
337	1.66082468638506e-06\\
338	1.30745773183505e-06\\
339	9.54090777285035e-07\\
340	1.41346781820005e-06\\
341	9.89427472740036e-07\\
342	9.89427472740036e-07\\
343	9.54090777285035e-07\\
344	1.02476416819504e-06\\
345	1.20144764547004e-06\\
346	1.09543755910504e-06\\
347	2.22621181366508e-06\\
348	2.15553842275508e-06\\
349	2.29688520457508e-06\\
350	1.83750816366007e-06\\
351	2.54424207276009e-06\\
352	2.29688520457508e-06\\
353	1.59015129547506e-06\\
354	1.87284485911507e-06\\
355	2.26154850912008e-06\\
356	3.03895580913011e-06\\
357	2.6149154636701e-06\\
358	3.07429250458511e-06\\
359	4.38175023642016e-06\\
360	3.85169980459514e-06\\
361	4.31107684551016e-06\\
362	5.5831978818902e-06\\
363	6.96132900463526e-06\\
364	5.22983092734019e-06\\
365	5.4771877955252e-06\\
366	4.80579058188018e-06\\
367	6.46661526826524e-06\\
368	5.65387127280021e-06\\
369	3.18030259095012e-06\\
370	3.35698606822512e-06\\
371	3.28631267731512e-06\\
372	2.33222190003009e-06\\
373	2.08486503184508e-06\\
374	1.90818155457007e-06\\
375	1.30745773183505e-06\\
376	1.44880451365505e-06\\
377	1.27212103638005e-06\\
378	1.09543755910504e-06\\
379	1.20144764547004e-06\\
380	1.59015129547506e-06\\
381	1.06010086365004e-06\\
382	1.13077425456004e-06\\
383	6.00723822735022e-07\\
384	1.09543755910504e-06\\
385	1.16611095001504e-06\\
386	1.44880451365505e-06\\
387	1.30745773183505e-06\\
388	1.83750816366007e-06\\
389	1.97885494548007e-06\\
390	1.90818155457007e-06\\
391	2.29688520457508e-06\\
392	3.10962920004011e-06\\
393	2.6855888545801e-06\\
394	1.80217146820507e-06\\
395	1.16611095001504e-06\\
396	1.02476416819504e-06\\
397	1.16611095001504e-06\\
398	1.94351825002507e-06\\
399	1.76683477275006e-06\\
400	3.03895580913011e-06\\
401	3.60434293641013e-06\\
402	2.7562622454901e-06\\
403	2.93294572276511e-06\\
404	4.02838328187015e-06\\
405	3.99304658641515e-06\\
406	3.14496589549512e-06\\
407	3.56900624095513e-06\\
408	3.10962920004011e-06\\
409	3.67501632732013e-06\\
410	3.85169980459514e-06\\
411	3.49833285004513e-06\\
412	3.92237319550514e-06\\
413	3.03895580913011e-06\\
414	3.53366954550013e-06\\
415	3.03895580913011e-06\\
416	2.8269356364001e-06\\
417	2.89760902731011e-06\\
418	3.56900624095513e-06\\
419	4.02838328187015e-06\\
420	7.31469595918527e-06\\
421	4.24040345460016e-06\\
422	2.33222190003009e-06\\
423	1.94351825002507e-06\\
424	2.57957876821509e-06\\
425	2.96828241822011e-06\\
426	3.74568971823014e-06\\
427	2.40289529094009e-06\\
428	1.06010086365004e-06\\
429	9.89427472740036e-07\\
430	7.77407300010029e-07\\
431	8.83417386375032e-07\\
432	7.77407300010029e-07\\
433	8.48080690920031e-07\\
434	7.77407300010029e-07\\
435	8.83417386375032e-07\\
436	8.1274399546503e-07\\
437	1.16611095001504e-06\\
438	9.18754081830034e-07\\
439	8.1274399546503e-07\\
440	7.77407300010029e-07\\
441	9.89427472740036e-07\\
442	8.83417386375032e-07\\
443	9.54090777285035e-07\\
444	1.02476416819504e-06\\
445	1.87284485911507e-06\\
446	1.55481460002006e-06\\
447	2.29688520457508e-06\\
448	2.93294572276511e-06\\
449	2.8269356364001e-06\\
450	2.12020172730008e-06\\
451	2.40289529094009e-06\\
452	2.01419164093507e-06\\
453	1.09543755910504e-06\\
454	1.83750816366007e-06\\
455	1.51947790456506e-06\\
456	2.01419164093507e-06\\
457	1.80217146820507e-06\\
458	2.50890537730509e-06\\
459	1.51947790456506e-06\\
460	1.69616138184006e-06\\
461	2.04952833639008e-06\\
462	2.22621181366508e-06\\
463	2.08486503184508e-06\\
464	2.08486503184508e-06\\
465	2.7915989409451e-06\\
466	3.35698606822512e-06\\
467	4.84112727733518e-06\\
468	4.38175023642016e-06\\
469	1.94351825002507e-06\\
470	1.02476416819504e-06\\
471	1.27212103638005e-06\\
472	1.27212103638005e-06\\
473	1.13077425456004e-06\\
474	8.83417386375032e-07\\
475	1.30745773183505e-06\\
476	8.48080690920031e-07\\
477	7.42070604555027e-07\\
478	8.48080690920031e-07\\
479	7.77407300010029e-07\\
480	7.06733909100026e-07\\
481	7.06733909100026e-07\\
482	9.54090777285035e-07\\
483	7.77407300010029e-07\\
484	6.71397213645025e-07\\
485	8.1274399546503e-07\\
486	1.20144764547004e-06\\
487	8.48080690920031e-07\\
488	6.36060518190023e-07\\
489	7.42070604555027e-07\\
490	8.1274399546503e-07\\
491	6.00723822735022e-07\\
492	8.1274399546503e-07\\
493	6.36060518190023e-07\\
494	7.42070604555027e-07\\
495	8.48080690920031e-07\\
496	8.48080690920031e-07\\
497	5.65387127280021e-07\\
498	7.42070604555027e-07\\
499	4.94713736370018e-07\\
500	7.06733909100026e-07\\
501	8.1274399546503e-07\\
502	8.1274399546503e-07\\
503	7.77407300010029e-07\\
504	1.06010086365004e-06\\
505	8.1274399546503e-07\\
506	1.51947790456506e-06\\
507	8.1274399546503e-07\\
508	1.02476416819504e-06\\
509	1.09543755910504e-06\\
510	1.06010086365004e-06\\
511	1.16611095001504e-06\\
512	7.77407300010029e-07\\
513	9.54090777285035e-07\\
514	9.18754081830034e-07\\
515	1.27212103638005e-06\\
516	6.00723822735022e-07\\
517	9.54090777285035e-07\\
518	8.83417386375032e-07\\
519	1.34279442729005e-06\\
520	1.02476416819504e-06\\
521	9.18754081830034e-07\\
522	1.13077425456004e-06\\
523	8.83417386375032e-07\\
524	1.02476416819504e-06\\
525	5.3005043182502e-07\\
526	5.65387127280021e-07\\
527	7.06733909100026e-07\\
528	5.65387127280021e-07\\
529	1.16611095001504e-06\\
530	6.36060518190023e-07\\
531	7.06733909100026e-07\\
532	7.77407300010029e-07\\
533	9.54090777285035e-07\\
534	6.00723822735022e-07\\
535	1.02476416819504e-06\\
536	8.1274399546503e-07\\
537	7.06733909100026e-07\\
538	7.42070604555027e-07\\
539	8.1274399546503e-07\\
540	4.59377040915017e-07\\
541	9.89427472740036e-07\\
542	8.1274399546503e-07\\
543	8.48080690920031e-07\\
544	6.00723822735022e-07\\
545	9.18754081830034e-07\\
546	7.77407300010029e-07\\
547	6.00723822735022e-07\\
548	8.1274399546503e-07\\
549	1.02476416819504e-06\\
550	8.48080690920031e-07\\
551	9.54090777285035e-07\\
552	9.18754081830034e-07\\
553	1.02476416819504e-06\\
554	1.16611095001504e-06\\
555	1.20144764547004e-06\\
556	1.27212103638005e-06\\
557	9.18754081830034e-07\\
558	1.37813112274505e-06\\
559	8.83417386375032e-07\\
560	9.18754081830034e-07\\
561	1.20144764547004e-06\\
562	9.54090777285035e-07\\
563	1.06010086365004e-06\\
564	7.42070604555027e-07\\
565	7.77407300010029e-07\\
566	9.89427472740036e-07\\
567	9.89427472740036e-07\\
568	8.83417386375032e-07\\
569	1.09543755910504e-06\\
570	1.34279442729005e-06\\
571	9.89427472740036e-07\\
572	9.18754081830034e-07\\
573	1.06010086365004e-06\\
574	8.83417386375032e-07\\
575	1.13077425456004e-06\\
576	8.83417386375032e-07\\
577	9.18754081830034e-07\\
578	9.54090777285035e-07\\
579	6.36060518190023e-07\\
580	7.77407300010029e-07\\
581	1.06010086365004e-06\\
582	8.48080690920031e-07\\
583	1.02476416819504e-06\\
584	1.16611095001504e-06\\
585	7.42070604555027e-07\\
586	8.48080690920031e-07\\
587	9.54090777285035e-07\\
588	1.06010086365004e-06\\
589	8.1274399546503e-07\\
590	9.18754081830034e-07\\
591	1.16611095001504e-06\\
592	1.06010086365004e-06\\
593	7.77407300010029e-07\\
594	8.83417386375032e-07\\
595	9.18754081830034e-07\\
596	4.59377040915017e-07\\
597	1.13077425456004e-06\\
598	1.02476416819504e-06\\
599	7.06733909100026e-07\\
600	1.02476416819504e-06\\
601	6.00723822735022e-07\\
602	8.83417386375032e-07\\
603	1.02476416819504e-06\\
604	1.30745773183505e-06\\
605	7.42070604555027e-07\\
606	1.16611095001504e-06\\
607	9.18754081830034e-07\\
608	1.06010086365004e-06\\
609	8.48080690920031e-07\\
610	8.83417386375032e-07\\
611	9.18754081830034e-07\\
612	7.06733909100026e-07\\
613	7.06733909100026e-07\\
614	7.77407300010029e-07\\
615	7.77407300010029e-07\\
616	7.42070604555027e-07\\
617	9.18754081830034e-07\\
618	1.13077425456004e-06\\
619	8.1274399546503e-07\\
620	6.36060518190023e-07\\
621	9.18754081830034e-07\\
622	8.83417386375032e-07\\
623	8.83417386375032e-07\\
624	6.00723822735022e-07\\
625	7.42070604555027e-07\\
626	9.18754081830034e-07\\
627	6.36060518190023e-07\\
628	6.00723822735022e-07\\
629	6.36060518190023e-07\\
630	6.71397213645025e-07\\
631	4.59377040915017e-07\\
632	3.53366954550013e-07\\
633	5.65387127280021e-07\\
634	8.48080690920031e-07\\
635	5.3005043182502e-07\\
636	8.1274399546503e-07\\
637	4.24040345460016e-07\\
638	4.59377040915017e-07\\
639	6.00723822735022e-07\\
640	5.3005043182502e-07\\
641	4.94713736370018e-07\\
642	5.65387127280021e-07\\
643	4.59377040915017e-07\\
644	6.00723822735022e-07\\
645	7.77407300010029e-07\\
646	5.3005043182502e-07\\
647	6.36060518190023e-07\\
648	7.77407300010029e-07\\
649	9.89427472740036e-07\\
650	3.88703650005014e-07\\
651	6.36060518190023e-07\\
652	4.94713736370018e-07\\
653	4.59377040915017e-07\\
654	6.36060518190023e-07\\
655	6.00723822735022e-07\\
656	6.36060518190023e-07\\
657	4.94713736370018e-07\\
658	4.94713736370018e-07\\
659	6.36060518190023e-07\\
660	8.48080690920031e-07\\
661	2.8269356364001e-07\\
662	3.53366954550013e-07\\
663	6.36060518190023e-07\\
664	1.09543755910504e-06\\
665	4.59377040915017e-07\\
666	4.24040345460016e-07\\
667	5.3005043182502e-07\\
668	4.94713736370018e-07\\
669	7.42070604555027e-07\\
670	5.3005043182502e-07\\
671	5.3005043182502e-07\\
672	2.8269356364001e-07\\
673	5.3005043182502e-07\\
674	5.65387127280021e-07\\
675	5.65387127280021e-07\\
676	7.06733909100026e-07\\
677	6.71397213645025e-07\\
678	7.06733909100026e-07\\
679	7.06733909100026e-07\\
680	6.36060518190023e-07\\
681	3.18030259095012e-07\\
682	4.59377040915017e-07\\
683	4.24040345460016e-07\\
684	4.94713736370018e-07\\
685	5.3005043182502e-07\\
686	4.24040345460016e-07\\
687	4.59377040915017e-07\\
688	2.8269356364001e-07\\
689	4.94713736370018e-07\\
690	3.18030259095012e-07\\
691	2.8269356364001e-07\\
692	4.59377040915017e-07\\
693	8.1274399546503e-07\\
694	4.59377040915017e-07\\
695	6.36060518190023e-07\\
696	4.24040345460016e-07\\
697	3.53366954550013e-07\\
698	2.8269356364001e-07\\
699	4.94713736370018e-07\\
700	6.36060518190023e-07\\
701	2.12020172730008e-07\\
702	3.53366954550013e-07\\
703	6.00723822735022e-07\\
704	3.88703650005014e-07\\
705	4.59377040915017e-07\\
706	4.94713736370018e-07\\
707	3.53366954550013e-07\\
708	4.94713736370018e-07\\
709	2.8269356364001e-07\\
710	2.8269356364001e-07\\
711	3.18030259095012e-07\\
712	1.76683477275006e-07\\
713	3.88703650005014e-07\\
714	1.76683477275006e-07\\
715	3.18030259095012e-07\\
716	1.76683477275006e-07\\
717	7.06733909100026e-08\\
718	3.53366954550013e-08\\
719	0\\
720	0\\
721	0\\
};
\draw[draw=black,very thick] (axis cs:50,0.0) rectangle (axis cs:721,0.05) node[fitting node] (zoom_rectangle) {};
\end{axis}

\begin{axis}[%
width=0.85\fwidth,
height=.6\fheight,
at={(\fwidth,1.2\fheight)},
anchor=north east,
scale only axis,
xmin=50,
xmax=721,
xlabel style={font=\color{white!15!black}},
ymin=0,
ymax=0.00005,
ylabel style={font=\color{white!15!black}},
ylabel={PDF},
axis background/.style={fill=white},
xmajorgrids,
ymajorgrids,
name=zoomed_in
]
\addplot[ybar interval, fill=mycolor1, fill opacity=0.6, draw=black, area legend] table[row sep=crcr] {%
x	y\\
41	0.000215271148711868\\
42	0.000170534892265836\\
43	0.000113536802496919\\
44	7.26169091600277e-05\\
45	3.66441531868363e-05\\
46	1.71736339911306e-05\\
47	6.07791161826022e-06\\
48	5.75988135916521e-06\\
49	4.55843371369517e-06\\
50	5.61853457734521e-06\\
51	5.15915753643019e-06\\
52	5.01781075461018e-06\\
53	4.94713736370018e-06\\
54	5.4771877955252e-06\\
55	5.26516762279519e-06\\
56	5.79521805462021e-06\\
57	5.61853457734521e-06\\
58	5.05314745006519e-06\\
59	5.3711777091602e-06\\
60	6.11324831371522e-06\\
61	4.98247405915518e-06\\
62	5.75988135916521e-06\\
63	4.91180066824518e-06\\
64	5.15915753643019e-06\\
65	5.5478611864352e-06\\
66	5.72454466371021e-06\\
67	6.43127857281024e-06\\
68	5.3358410137052e-06\\
69	4.80579058188018e-06\\
70	5.83055475007521e-06\\
71	6.36060518190023e-06\\
72	6.53728865917524e-06\\
73	9.22287751375534e-06\\
74	1.37106378365405e-05\\
75	2.08839870139058e-05\\
76	1.27565470592555e-05\\
77	2.14847108366408e-05\\
78	2.01065797138957e-05\\
79	2.6325838113976e-05\\
80	4.18386474187215e-05\\
81	5.08848414552019e-05\\
82	3.92590686505064e-05\\
83	2.6255164723066e-05\\
84	2.41349629957659e-05\\
85	2.8446039841276e-05\\
86	2.58664610730609e-05\\
87	2.7809979323086e-05\\
88	2.39582795184909e-05\\
89	2.18380777911908e-05\\
90	1.57248294774756e-05\\
91	1.22971700183405e-05\\
92	8.1274399546503e-06\\
93	6.71397213645025e-06\\
94	6.43127857281024e-06\\
95	4.55843371369517e-06\\
96	4.16973006369015e-06\\
97	4.80579058188018e-06\\
98	5.93656483644022e-06\\
99	4.77045388642518e-06\\
100	5.26516762279519e-06\\
101	5.5125244909802e-06\\
102	4.13439336823515e-06\\
103	5.5478611864352e-06\\
104	4.62910710460517e-06\\
105	6.14858500917023e-06\\
106	6.92599230918025e-06\\
107	7.56205282737028e-06\\
108	7.52671613191528e-06\\
109	9.54090777285035e-06\\
110	1.16611095001504e-05\\
111	1.12724058501454e-05\\
112	9.71759125012536e-06\\
113	1.26858736683455e-05\\
114	1.80217146820507e-05\\
115	1.73149807729506e-05\\
116	2.04246099729908e-05\\
117	2.8163346277636e-05\\
118	3.94357521277814e-05\\
119	5.07434946733819e-05\\
120	2.53010739457809e-05\\
121	4.23686978505466e-05\\
122	3.66441531868363e-05\\
123	1.98592228457107e-05\\
124	1.82690715502357e-05\\
125	2.24388016139258e-05\\
126	1.72796440774956e-05\\
127	1.58661762592956e-05\\
128	1.62548799093006e-05\\
129	2.22974548321058e-05\\
130	2.91174370549211e-05\\
131	3.24744231231462e-05\\
132	3.56547257140963e-05\\
133	3.29338001640612e-05\\
134	2.91174370549211e-05\\
135	2.26508217866558e-05\\
136	1.89758054593357e-05\\
137	1.79157045956857e-05\\
138	1.60075230411156e-05\\
139	1.31805874047155e-05\\
140	1.23325067137955e-05\\
141	1.14844260228754e-05\\
142	1.42760249638205e-05\\
143	8.69282708193032e-06\\
144	8.0567665637403e-06\\
145	8.79883716829532e-06\\
146	9.61158116376035e-06\\
147	1.42760249638205e-05\\
148	9.64691785921536e-06\\
149	7.73873630464528e-06\\
150	8.0214298682853e-06\\
151	8.37479682283531e-06\\
152	1.00356215092204e-05\\
153	1.62195432138456e-05\\
154	2.6290501418521e-05\\
155	2.19440878775558e-05\\
156	3.13083121731311e-05\\
157	4.15559538550815e-05\\
158	2.00005696275307e-05\\
159	1.82690715502357e-05\\
160	3.00715278322061e-05\\
161	3.12023020867661e-05\\
162	9.68225455467036e-06\\
163	9.43489768648535e-06\\
164	2.8587386623096e-05\\
165	2.31455355230258e-05\\
166	1.10250489819604e-05\\
167	1.16611095001504e-05\\
168	1.60428597365706e-05\\
169	1.14490893274204e-05\\
170	1.05303352455904e-05\\
171	1.26858736683455e-05\\
172	1.40993414865455e-05\\
173	1.97885494548007e-05\\
174	3.14143222594962e-05\\
175	3.10962920004011e-05\\
176	3.59020825822813e-05\\
177	4.18033107232665e-05\\
178	3.89763750868664e-05\\
179	3.32164937277012e-05\\
180	3.08135984367611e-05\\
181	2.8587386623096e-05\\
182	2.25094750048358e-05\\
183	1.83750816366007e-05\\
184	1.08483655046854e-05\\
185	6.78464552736025e-06\\
186	5.4418511000702e-06\\
187	4.62910710460517e-06\\
188	4.55843371369517e-06\\
189	4.91180066824518e-06\\
190	3.92237319550514e-06\\
191	3.56900624095513e-06\\
192	3.46299615459013e-06\\
193	3.42765945913513e-06\\
194	3.46299615459013e-06\\
195	3.88703650005014e-06\\
196	3.25097598186012e-06\\
197	3.60434293641013e-06\\
198	3.92237319550514e-06\\
199	3.92237319550514e-06\\
200	3.00361911367511e-06\\
201	3.63967963186513e-06\\
202	3.71035302277514e-06\\
203	3.99304658641515e-06\\
204	3.88703650005014e-06\\
205	4.20506675914515e-06\\
206	3.92237319550514e-06\\
207	4.52309701824017e-06\\
208	4.48776032278517e-06\\
209	4.62910710460517e-06\\
210	5.19449423188519e-06\\
211	5.15915753643019e-06\\
212	6.07791161826022e-06\\
213	8.94018395011533e-06\\
214	1.13430792410554e-05\\
215	2.05306200593558e-05\\
216	1.93291724138857e-05\\
217	2.55837675094209e-05\\
218	2.33222190003009e-05\\
219	3.40999111140763e-05\\
220	2.90821003594661e-05\\
221	1.40286680956355e-05\\
222	1.57955028683856e-05\\
223	1.80923880729607e-05\\
224	1.25445268865255e-05\\
225	1.26505369728905e-05\\
226	1.22618333228854e-05\\
227	1.13784159365104e-05\\
228	1.51241056547406e-05\\
229	2.8269356364001e-05\\
230	2.00005696275307e-05\\
231	1.20144764547004e-05\\
232	8.1627766501053e-06\\
233	6.71397213645025e-06\\
234	4.73511719097017e-06\\
235	3.60434293641013e-06\\
236	4.27574015005516e-06\\
237	3.81636310914014e-06\\
238	4.20506675914515e-06\\
239	3.46299615459013e-06\\
240	3.46299615459013e-06\\
241	3.56900624095513e-06\\
242	3.56900624095513e-06\\
243	3.07429250458511e-06\\
244	3.42765945913513e-06\\
245	4.45242362733016e-06\\
246	5.08848414552019e-06\\
247	6.64329874554024e-06\\
248	5.83055475007521e-06\\
249	6.57262535463024e-06\\
250	9.01085734102533e-06\\
251	1.69969505138556e-05\\
252	1.62902166047556e-05\\
253	1.24385168001605e-05\\
254	1.15197627183304e-05\\
255	1.41700148774555e-05\\
256	1.13077425456004e-05\\
257	9.15220412284534e-06\\
258	1.04596618546804e-05\\
259	1.22618333228854e-05\\
260	2.27921685684758e-05\\
261	3.19090359958662e-05\\
262	9.04619403648033e-06\\
263	3.49833285004513e-06\\
264	5.08848414552019e-06\\
265	6.11324831371522e-06\\
266	5.61853457734521e-06\\
267	1.02123049864954e-05\\
268	2.45590033412259e-05\\
269	6.28993179099023e-06\\
270	6.39594187735523e-06\\
271	7.56205282737028e-06\\
272	8.51614360465531e-06\\
273	7.84474639101029e-06\\
274	9.43489768648535e-06\\
275	8.79883716829532e-06\\
276	8.51614360465531e-06\\
277	7.91541978192029e-06\\
278	6.60796205008524e-06\\
279	4.69978049551517e-06\\
280	4.16973006369015e-06\\
281	2.8622723318551e-06\\
282	2.47356868185009e-06\\
283	2.50890537730509e-06\\
284	3.18030259095012e-06\\
285	2.54424207276009e-06\\
286	1.59015129547506e-06\\
287	2.04952833639008e-06\\
288	1.87284485911507e-06\\
289	1.83750816366007e-06\\
290	1.97885494548007e-06\\
291	2.22621181366508e-06\\
292	2.15553842275508e-06\\
293	2.57957876821509e-06\\
294	2.7915989409451e-06\\
295	3.67501632732013e-06\\
296	3.63967963186513e-06\\
297	3.49833285004513e-06\\
298	3.07429250458511e-06\\
299	4.09905667278015e-06\\
300	4.94713736370018e-06\\
301	4.66444380006017e-06\\
302	2.8269356364001e-06\\
303	2.93294572276511e-06\\
304	3.63967963186513e-06\\
305	2.89760902731011e-06\\
306	2.43823198639509e-06\\
307	2.6149154636701e-06\\
308	3.14496589549512e-06\\
309	3.49833285004513e-06\\
310	3.60434293641013e-06\\
311	3.95770989096015e-06\\
312	4.45242362733016e-06\\
313	4.24040345460016e-06\\
314	3.92237319550514e-06\\
315	4.24040345460016e-06\\
316	6.11324831371522e-06\\
317	6.46661526826524e-06\\
318	6.74930883190525e-06\\
319	7.59738952282528e-06\\
320	6.99666570009026e-06\\
321	6.81998222281525e-06\\
322	8.2334500410153e-06\\
323	7.38536935009527e-06\\
324	9.75292794558036e-06\\
325	7.66806291373528e-06\\
326	4.27574015005516e-06\\
327	3.46299615459013e-06\\
328	2.54424207276009e-06\\
329	1.90818155457007e-06\\
330	1.48414120911005e-06\\
331	2.22621181366508e-06\\
332	2.19087511821008e-06\\
333	1.55481460002006e-06\\
334	1.76683477275006e-06\\
335	2.15553842275508e-06\\
336	1.51947790456506e-06\\
337	1.66082468638506e-06\\
338	1.30745773183505e-06\\
339	9.54090777285035e-07\\
340	1.41346781820005e-06\\
341	9.89427472740036e-07\\
342	9.89427472740036e-07\\
343	9.54090777285035e-07\\
344	1.02476416819504e-06\\
345	1.20144764547004e-06\\
346	1.09543755910504e-06\\
347	2.22621181366508e-06\\
348	2.15553842275508e-06\\
349	2.29688520457508e-06\\
350	1.83750816366007e-06\\
351	2.54424207276009e-06\\
352	2.29688520457508e-06\\
353	1.59015129547506e-06\\
354	1.87284485911507e-06\\
355	2.26154850912008e-06\\
356	3.03895580913011e-06\\
357	2.6149154636701e-06\\
358	3.07429250458511e-06\\
359	4.38175023642016e-06\\
360	3.85169980459514e-06\\
361	4.31107684551016e-06\\
362	5.5831978818902e-06\\
363	6.96132900463526e-06\\
364	5.22983092734019e-06\\
365	5.4771877955252e-06\\
366	4.80579058188018e-06\\
367	6.46661526826524e-06\\
368	5.65387127280021e-06\\
369	3.18030259095012e-06\\
370	3.35698606822512e-06\\
371	3.28631267731512e-06\\
372	2.33222190003009e-06\\
373	2.08486503184508e-06\\
374	1.90818155457007e-06\\
375	1.30745773183505e-06\\
376	1.44880451365505e-06\\
377	1.27212103638005e-06\\
378	1.09543755910504e-06\\
379	1.20144764547004e-06\\
380	1.59015129547506e-06\\
381	1.06010086365004e-06\\
382	1.13077425456004e-06\\
383	6.00723822735022e-07\\
384	1.09543755910504e-06\\
385	1.16611095001504e-06\\
386	1.44880451365505e-06\\
387	1.30745773183505e-06\\
388	1.83750816366007e-06\\
389	1.97885494548007e-06\\
390	1.90818155457007e-06\\
391	2.29688520457508e-06\\
392	3.10962920004011e-06\\
393	2.6855888545801e-06\\
394	1.80217146820507e-06\\
395	1.16611095001504e-06\\
396	1.02476416819504e-06\\
397	1.16611095001504e-06\\
398	1.94351825002507e-06\\
399	1.76683477275006e-06\\
400	3.03895580913011e-06\\
401	3.60434293641013e-06\\
402	2.7562622454901e-06\\
403	2.93294572276511e-06\\
404	4.02838328187015e-06\\
405	3.99304658641515e-06\\
406	3.14496589549512e-06\\
407	3.56900624095513e-06\\
408	3.10962920004011e-06\\
409	3.67501632732013e-06\\
410	3.85169980459514e-06\\
411	3.49833285004513e-06\\
412	3.92237319550514e-06\\
413	3.03895580913011e-06\\
414	3.53366954550013e-06\\
415	3.03895580913011e-06\\
416	2.8269356364001e-06\\
417	2.89760902731011e-06\\
418	3.56900624095513e-06\\
419	4.02838328187015e-06\\
420	7.31469595918527e-06\\
421	4.24040345460016e-06\\
422	2.33222190003009e-06\\
423	1.94351825002507e-06\\
424	2.57957876821509e-06\\
425	2.96828241822011e-06\\
426	3.74568971823014e-06\\
427	2.40289529094009e-06\\
428	1.06010086365004e-06\\
429	9.89427472740036e-07\\
430	7.77407300010029e-07\\
431	8.83417386375032e-07\\
432	7.77407300010029e-07\\
433	8.48080690920031e-07\\
434	7.77407300010029e-07\\
435	8.83417386375032e-07\\
436	8.1274399546503e-07\\
437	1.16611095001504e-06\\
438	9.18754081830034e-07\\
439	8.1274399546503e-07\\
440	7.77407300010029e-07\\
441	9.89427472740036e-07\\
442	8.83417386375032e-07\\
443	9.54090777285035e-07\\
444	1.02476416819504e-06\\
445	1.87284485911507e-06\\
446	1.55481460002006e-06\\
447	2.29688520457508e-06\\
448	2.93294572276511e-06\\
449	2.8269356364001e-06\\
450	2.12020172730008e-06\\
451	2.40289529094009e-06\\
452	2.01419164093507e-06\\
453	1.09543755910504e-06\\
454	1.83750816366007e-06\\
455	1.51947790456506e-06\\
456	2.01419164093507e-06\\
457	1.80217146820507e-06\\
458	2.50890537730509e-06\\
459	1.51947790456506e-06\\
460	1.69616138184006e-06\\
461	2.04952833639008e-06\\
462	2.22621181366508e-06\\
463	2.08486503184508e-06\\
464	2.08486503184508e-06\\
465	2.7915989409451e-06\\
466	3.35698606822512e-06\\
467	4.84112727733518e-06\\
468	4.38175023642016e-06\\
469	1.94351825002507e-06\\
470	1.02476416819504e-06\\
471	1.27212103638005e-06\\
472	1.27212103638005e-06\\
473	1.13077425456004e-06\\
474	8.83417386375032e-07\\
475	1.30745773183505e-06\\
476	8.48080690920031e-07\\
477	7.42070604555027e-07\\
478	8.48080690920031e-07\\
479	7.77407300010029e-07\\
480	7.06733909100026e-07\\
481	7.06733909100026e-07\\
482	9.54090777285035e-07\\
483	7.77407300010029e-07\\
484	6.71397213645025e-07\\
485	8.1274399546503e-07\\
486	1.20144764547004e-06\\
487	8.48080690920031e-07\\
488	6.36060518190023e-07\\
489	7.42070604555027e-07\\
490	8.1274399546503e-07\\
491	6.00723822735022e-07\\
492	8.1274399546503e-07\\
493	6.36060518190023e-07\\
494	7.42070604555027e-07\\
495	8.48080690920031e-07\\
496	8.48080690920031e-07\\
497	5.65387127280021e-07\\
498	7.42070604555027e-07\\
499	4.94713736370018e-07\\
500	7.06733909100026e-07\\
501	8.1274399546503e-07\\
502	8.1274399546503e-07\\
503	7.77407300010029e-07\\
504	1.06010086365004e-06\\
505	8.1274399546503e-07\\
506	1.51947790456506e-06\\
507	8.1274399546503e-07\\
508	1.02476416819504e-06\\
509	1.09543755910504e-06\\
510	1.06010086365004e-06\\
511	1.16611095001504e-06\\
512	7.77407300010029e-07\\
513	9.54090777285035e-07\\
514	9.18754081830034e-07\\
515	1.27212103638005e-06\\
516	6.00723822735022e-07\\
517	9.54090777285035e-07\\
518	8.83417386375032e-07\\
519	1.34279442729005e-06\\
520	1.02476416819504e-06\\
521	9.18754081830034e-07\\
522	1.13077425456004e-06\\
523	8.83417386375032e-07\\
524	1.02476416819504e-06\\
525	5.3005043182502e-07\\
526	5.65387127280021e-07\\
527	7.06733909100026e-07\\
528	5.65387127280021e-07\\
529	1.16611095001504e-06\\
530	6.36060518190023e-07\\
531	7.06733909100026e-07\\
532	7.77407300010029e-07\\
533	9.54090777285035e-07\\
534	6.00723822735022e-07\\
535	1.02476416819504e-06\\
536	8.1274399546503e-07\\
537	7.06733909100026e-07\\
538	7.42070604555027e-07\\
539	8.1274399546503e-07\\
540	4.59377040915017e-07\\
541	9.89427472740036e-07\\
542	8.1274399546503e-07\\
543	8.48080690920031e-07\\
544	6.00723822735022e-07\\
545	9.18754081830034e-07\\
546	7.77407300010029e-07\\
547	6.00723822735022e-07\\
548	8.1274399546503e-07\\
549	1.02476416819504e-06\\
550	8.48080690920031e-07\\
551	9.54090777285035e-07\\
552	9.18754081830034e-07\\
553	1.02476416819504e-06\\
554	1.16611095001504e-06\\
555	1.20144764547004e-06\\
556	1.27212103638005e-06\\
557	9.18754081830034e-07\\
558	1.37813112274505e-06\\
559	8.83417386375032e-07\\
560	9.18754081830034e-07\\
561	1.20144764547004e-06\\
562	9.54090777285035e-07\\
563	1.06010086365004e-06\\
564	7.42070604555027e-07\\
565	7.77407300010029e-07\\
566	9.89427472740036e-07\\
567	9.89427472740036e-07\\
568	8.83417386375032e-07\\
569	1.09543755910504e-06\\
570	1.34279442729005e-06\\
571	9.89427472740036e-07\\
572	9.18754081830034e-07\\
573	1.06010086365004e-06\\
574	8.83417386375032e-07\\
575	1.13077425456004e-06\\
576	8.83417386375032e-07\\
577	9.18754081830034e-07\\
578	9.54090777285035e-07\\
579	6.36060518190023e-07\\
580	7.77407300010029e-07\\
581	1.06010086365004e-06\\
582	8.48080690920031e-07\\
583	1.02476416819504e-06\\
584	1.16611095001504e-06\\
585	7.42070604555027e-07\\
586	8.48080690920031e-07\\
587	9.54090777285035e-07\\
588	1.06010086365004e-06\\
589	8.1274399546503e-07\\
590	9.18754081830034e-07\\
591	1.16611095001504e-06\\
592	1.06010086365004e-06\\
593	7.77407300010029e-07\\
594	8.83417386375032e-07\\
595	9.18754081830034e-07\\
596	4.59377040915017e-07\\
597	1.13077425456004e-06\\
598	1.02476416819504e-06\\
599	7.06733909100026e-07\\
600	1.02476416819504e-06\\
601	6.00723822735022e-07\\
602	8.83417386375032e-07\\
603	1.02476416819504e-06\\
604	1.30745773183505e-06\\
605	7.42070604555027e-07\\
606	1.16611095001504e-06\\
607	9.18754081830034e-07\\
608	1.06010086365004e-06\\
609	8.48080690920031e-07\\
610	8.83417386375032e-07\\
611	9.18754081830034e-07\\
612	7.06733909100026e-07\\
613	7.06733909100026e-07\\
614	7.77407300010029e-07\\
615	7.77407300010029e-07\\
616	7.42070604555027e-07\\
617	9.18754081830034e-07\\
618	1.13077425456004e-06\\
619	8.1274399546503e-07\\
620	6.36060518190023e-07\\
621	9.18754081830034e-07\\
622	8.83417386375032e-07\\
623	8.83417386375032e-07\\
624	6.00723822735022e-07\\
625	7.42070604555027e-07\\
626	9.18754081830034e-07\\
627	6.36060518190023e-07\\
628	6.00723822735022e-07\\
629	6.36060518190023e-07\\
630	6.71397213645025e-07\\
631	4.59377040915017e-07\\
632	3.53366954550013e-07\\
633	5.65387127280021e-07\\
634	8.48080690920031e-07\\
635	5.3005043182502e-07\\
636	8.1274399546503e-07\\
637	4.24040345460016e-07\\
638	4.59377040915017e-07\\
639	6.00723822735022e-07\\
640	5.3005043182502e-07\\
641	4.94713736370018e-07\\
642	5.65387127280021e-07\\
643	4.59377040915017e-07\\
644	6.00723822735022e-07\\
645	7.77407300010029e-07\\
646	5.3005043182502e-07\\
647	6.36060518190023e-07\\
648	7.77407300010029e-07\\
649	9.89427472740036e-07\\
650	3.88703650005014e-07\\
651	6.36060518190023e-07\\
652	4.94713736370018e-07\\
653	4.59377040915017e-07\\
654	6.36060518190023e-07\\
655	6.00723822735022e-07\\
656	6.36060518190023e-07\\
657	4.94713736370018e-07\\
658	4.94713736370018e-07\\
659	6.36060518190023e-07\\
660	8.48080690920031e-07\\
661	2.8269356364001e-07\\
662	3.53366954550013e-07\\
663	6.36060518190023e-07\\
664	1.09543755910504e-06\\
665	4.59377040915017e-07\\
666	4.24040345460016e-07\\
667	5.3005043182502e-07\\
668	4.94713736370018e-07\\
669	7.42070604555027e-07\\
670	5.3005043182502e-07\\
671	5.3005043182502e-07\\
672	2.8269356364001e-07\\
673	5.3005043182502e-07\\
674	5.65387127280021e-07\\
675	5.65387127280021e-07\\
676	7.06733909100026e-07\\
677	6.71397213645025e-07\\
678	7.06733909100026e-07\\
679	7.06733909100026e-07\\
680	6.36060518190023e-07\\
681	3.18030259095012e-07\\
682	4.59377040915017e-07\\
683	4.24040345460016e-07\\
684	4.94713736370018e-07\\
685	5.3005043182502e-07\\
686	4.24040345460016e-07\\
687	4.59377040915017e-07\\
688	2.8269356364001e-07\\
689	4.94713736370018e-07\\
690	3.18030259095012e-07\\
691	2.8269356364001e-07\\
692	4.59377040915017e-07\\
693	8.1274399546503e-07\\
694	4.59377040915017e-07\\
695	6.36060518190023e-07\\
696	4.24040345460016e-07\\
697	3.53366954550013e-07\\
698	2.8269356364001e-07\\
699	4.94713736370018e-07\\
700	6.36060518190023e-07\\
701	2.12020172730008e-07\\
702	3.53366954550013e-07\\
703	6.00723822735022e-07\\
704	3.88703650005014e-07\\
705	4.59377040915017e-07\\
706	4.94713736370018e-07\\
707	3.53366954550013e-07\\
708	4.94713736370018e-07\\
709	2.8269356364001e-07\\
710	2.8269356364001e-07\\
711	3.18030259095012e-07\\
712	1.76683477275006e-07\\
713	3.88703650005014e-07\\
714	1.76683477275006e-07\\
715	3.18030259095012e-07\\
716	1.76683477275006e-07\\
717	7.06733909100026e-08\\
718	3.53366954550013e-08\\
719	0\\
720	0\\
721	0\\
};

\end{axis}

\draw[draw=black] (zoom_rectangle.north east) -- (zoomed_in.south east);
\draw[draw=black] (zoom_rectangle.north west) -- (zoomed_in.south west);

\end{tikzpicture}%

%% file: figures/latency_g2s.tikz
%
%
\definecolor{mycolor1}{rgb}{0.00000,0.44700,0.74100}%
\definecolor{mycolor2}{rgb}{0.85000,0.32500,0.09800}%
\definecolor{mycolor3}{rgb}{0.92900,0.69400,0.12500}%
\definecolor{mycolor4}{rgb}{0.4940,0.1840,0.5560}
\begin{tikzpicture}

\begin{axis}[%
width=0.951\fwidth,
height=\fheight,
at={(0\fwidth,0\fheight)},
scale only axis,
xmin=0,
xmax=40,
xlabel style={font=\color{white!15!black}},
xlabel={$X =$ Round-Trip Time [ms]},
ymin=0,
ymax=1,
ylabel style={font=\color{white!15!black}},
ylabel={$F(X)$},
axis background/.style={fill=white},
xmajorgrids,
ymajorgrids,
legend style={anchor=south, at={(0.5\fwidth,1.\fheight)}, align=left, draw=white!15!black,},
legend columns=2,
]
\addplot [color=mycolor1]
  table[row sep=crcr]{%
1.20083222883862	0\\
1.22549759431584	0.00184714921387297\\
1.30362295054242	0.00790592254571365\\
1.32950675912982	0.00998496111945535\\
1.37028258911837	0.0133430789803306\\
1.43618912957202	0.0189941282723165\\
1.46892277385889	0.0218453938173671\\
1.55700372014006	0.0295598453776744\\
1.6308564213124	0.0361893438579521\\
1.6543795208822	0.0383431711756188\\
1.82751025011296	0.0543318835333242\\
1.90438656567436	0.0618331797978335\\
1.95925924105439	0.0672582402226247\\
1.97262498039436	0.0685967669728829\\
2.0416602057093	0.0755354512663526\\
2.09282173972044	0.0807836444249315\\
2.11098451594728	0.0826566082940143\\
2.19322127055841	0.0910972629832667\\
2.22763745886533	0.0946394762138354\\
2.24850478085935	0.0968228059946092\\
2.35954653293465	0.108659784234449\\
2.39882639753277	0.112941919239239\\
2.41449695899619	0.114672091186991\\
2.45253921697669	0.118843444442907\\
2.47851492396734	0.121717721909191\\
2.52289706512596	0.126619261126516\\
2.65098507452617	0.140911729369195\\
2.6809656380151	0.144342570801625\\
2.79099169008554	0.157018599091757\\
2.81225727252817	0.159523063183256\\
3.12183102844481	0.195952262066761\\
3.16312964987727	0.200930655200469\\
3.2301346216315	0.208899417992757\\
3.40088847135198	0.229054615700363\\
3.4378516577656	0.233532499547835\\
3.55588969761312	0.247804168554225\\
3.57024805803844	0.249511033556171\\
3.58244093747166	0.250998547745048\\
3.74877097411143	0.270925543900772\\
3.84211181825463	0.282141937829818\\
3.88565733808131	0.287396621530394\\
3.90178212495846	0.289349094537553\\
3.93135017035939	0.292947214935767\\
4.02934988403587	0.304836265023251\\
4.04292571836489	0.306494745985741\\
4.55930802221882	0.369178629308982\\
4.58421405801346	0.372168703942952\\
4.63794984044076	0.378615139640043\\
4.67246522904403	0.382767758832024\\
5.07977706156509	0.431241043247649\\
5.14761374111939	0.439165552345488\\
5.23122101107899	0.448939718366661\\
5.44107037272263	0.473286035981092\\
5.48092771083579	0.477870276208082\\
5.55755434295045	0.486798164075596\\
5.75563113351533	0.509692075431399\\
5.79613288933575	0.514348596692567\\
6.01257091918643	0.539321694133177\\
6.03258496733878	0.541605627312967\\
6.05124046016604	0.543731427290615\\
6.08000965169058	0.547033933008425\\
6.13894044032039	0.553826137572331\\
6.32170702710878	0.574660186954116\\
6.4008447320118	0.583640884229656\\
6.43385191396376	0.587424280095734\\
6.45116826340824	0.589386931452488\\
6.54200532018381	0.599651280896836\\
6.68377790856675	0.615803879436029\\
6.86683894783423	0.636532014975014\\
6.93889907533296	0.644684283086995\\
6.96901646970107	0.648086064593016\\
7.03841018194584	0.655852292975515\\
7.05603512680125	0.657833530883984\\
7.11904749453972	0.664907041433221\\
7.52878138066067	0.711096392663343\\
7.64578406232861	0.724462041026884\\
7.68118975431424	0.728498568026115\\
7.71433448203373	0.732212780616145\\
7.81026874804734	0.74301392737603\\
7.96915228635473	0.76079004645921\\
7.98845318170418	0.76295832998378\\
8.03095396354868	0.767712061860969\\
8.06558182349647	0.771564198466221\\
8.10178451787731	0.775585679209595\\
8.11782524101486	0.777386067019476\\
8.2027498841465	0.786808416167856\\
8.61364179966434	0.832160192474413\\
8.63512853007215	0.834513751467179\\
8.65354427820516	0.83650959309502\\
8.68367898231423	0.83977758093117\\
8.71008226893342	0.842653923569843\\
8.7625741998484	0.848407346408768\\
9.07339471513524	0.882333703835647\\
9.0919105722348	0.884353294946244\\
9.23108727583854	0.899402206342085\\
9.27224564634811	0.903821527800128\\
9.67767553871391	0.947249448488156\\
9.73200719660209	0.953034291450248\\
9.80338788152599	0.960652122443925\\
9.88392220264714	0.969168745975546\\
9.91224513750272	0.972183455166094\\
9.97050693451298	0.97833339110891\\
10.0162591166836	0.983209853232578\\
10.0619193976252	0.988063155922703\\
10.1744366287974	1\\
};
\addlegendentry{Lower LEO}

\addplot [color=mycolor3]
  table[row sep=crcr]{%
1.38571633393309	0\\
2.6668517747075	0.000309775862415762\\
3.06053672063976	0.0010325862080407\\
3.3269636834933	0.00256671428856947\\
3.67330790871915	0.0058709901542997\\
3.80783298661557	0.00890974385225007\\
3.87543036558941	0.0108421551844451\\
3.93670129838081	0.0126860591273861\\
4.07424919580421	0.0169344138118959\\
4.12614300766412	0.0185570492816858\\
4.34023724721791	0.025549133033298\\
4.43644490717285	0.0288239064359601\\
4.48481484202205	0.0304760443688394\\
5.0564902082362	0.0506999828149937\\
5.37191262584679	0.0626927340598762\\
5.39793404622023	0.0636958178048452\\
5.67352429442693	0.0749067537779631\\
5.81669037401135	0.0809105050161989\\
6.17282804160671	0.09628128828464\\
6.35540890238609	0.104586231643687\\
6.39648832078291	0.106459638049724\\
6.85489961379442	0.128306211965842\\
7.16004937802529	0.143854010012898\\
7.34984391928264	0.153840593767988\\
7.41832417631852	0.157484147959273\\
7.47207793123009	0.160390140573387\\
7.50272252370928	0.16208653220091\\
7.58942325830944	0.166836428757996\\
7.66870328835682	0.171261798221096\\
7.79433969473171	0.178342389362104\\
7.89257714301279	0.183977359811827\\
8.18408421085895	0.201059285939532\\
8.25583987403596	0.205337143087235\\
8.27295522313585	0.206369729295293\\
8.52806937259613	0.221725761332429\\
8.55772358482986	0.22355491404387\\
8.734844452228	0.234633088933315\\
8.81401740793266	0.239619005195138\\
8.85832994244101	0.242451241651576\\
9.09047178497106	0.257202473195449\\
9.31421988153237	0.271894699813199\\
9.53074588406038	0.286542672736317\\
9.56339866340608	0.288740606236356\\
9.65395849793694	0.294906621021749\\
9.84118269408659	0.307695938770365\\
9.87818582116131	0.310277404290556\\
9.93734972748501	0.314392997891332\\
9.98238579363435	0.317549761441754\\
10.0479841380479	0.322196399378097\\
10.1380502263925	0.328716443720563\\
10.3919638449885	0.347376751623777\\
10.6686103687899	0.367954719627768\\
10.7047894777003	0.370639443768781\\
10.7255810570056	0.372203074312448\\
10.7674024590868	0.375300832936713\\
10.8415436790159	0.380876798460385\\
10.976516323156	0.391158406846625\\
11.0333162902613	0.395539522615245\\
11.153273075484	0.404788544793433\\
11.2900471314162	0.415320924115964\\
11.662311513185	0.444380850257932\\
11.7778452138834	0.453644623667682\\
11.8175943050267	0.456801387218139\\
11.9275072511858	0.465622623681583\\
11.958495689663	0.468115581812548\\
11.9849788906174	0.470284012849536\\
12.0361406059989	0.474399606450362\\
12.1491051318458	0.483515867544693\\
12.218222151006	0.489106584299943\\
12.3956708220129	0.503562791212964\\
12.4574060838106	0.508637214863757\\
12.5696334358922	0.517959993198918\\
12.5889652265994	0.519582628668648\\
12.8019273238551	0.53766763854027\\
12.8860407457367	0.544895741996292\\
12.9175880756908	0.547595217368634\\
12.9755655640948	0.552595884861656\\
13.0113355401604	0.555664141022564\\
13.0290471985406	0.557198269103022\\
13.0560828162072	0.559528963686788\\
13.0749116407132	0.561122096693413\\
13.2829683228602	0.579177604101766\\
13.3573318987287	0.585682897212113\\
13.5773242356291	0.605110269153826\\
13.6234180611102	0.609211111522686\\
13.6965577684857	0.615775409559152\\
14.1957431035638	0.661076441626278\\
14.268235223025	0.667729247051916\\
14.2955696776113	0.670236956414129\\
14.4314393923921	0.682775503225173\\
15.252338423509	0.758788599363399\\
15.3633915623607	0.769335729916953\\
15.3949532620536	0.772344981151814\\
15.4211875500361	0.774867441745744\\
15.4667163745803	0.779233806282608\\
15.5008534811222	0.782508579685256\\
15.9594983631414	0.826422995990203\\
15.9818854018918	0.82856192456401\\
16.0124542515048	0.831497419641163\\
16.0553848459536	0.835613013241804\\
16.1800447037009	0.847473003402801\\
16.2245158302697	0.851514840845724\\
16.3487280243622	0.862475005882569\\
16.6232648284747	0.885899961573564\\
16.9661760943389	0.913809291653411\\
17.005542782639	0.916862796582855\\
17.1308855833096	0.926362589696684\\
17.1750739590311	0.929711119256989\\
17.2396470114149	0.934520020740077\\
17.2999555978095	0.939033897592328\\
17.4235761441448	0.94817966114929\\
17.4997144738114	0.953814631598899\\
17.8995608862856	0.982771299118788\\
17.9583923274977	0.98631159468923\\
18.0786634417824	0.991917062675753\\
18.1175655470441	0.993436439524732\\
18.2186655408849	0.996622705538119\\
18.2613174235902	0.997478276967641\\
18.3060045170604	0.998112579924012\\
18.3759772839911	0.998820639038097\\
18.527472970913	0.999690961699166\\
18.6546269432894	0.999971235098489\\
18.6933880442883	0.999985986330035\\
};
\addlegendentry{Starlink}

\addplot [color=mycolor2]
  table[row sep=crcr]{%
13.342576730321	0\\
13.5094958345453	0.00144473561736902\\
13.6629475641024	0.00275803776168715\\
13.7948188835902	0.00389742288609796\\
13.9029642546891	0.00488767305961346\\
13.9874324286951	0.00569014005557733\\
14.4687571119001	0.0104016834105565\\
15.2451877928492	0.0186647332719474\\
15.6818880147009	0.023680594533694\\
15.8708696141444	0.0258987372208779\\
16.0272395090236	0.0277393958928798\\
16.1048432591746	0.0286734438742187\\
16.2654846177388	0.0306042325709512\\
16.4579035302219	0.0329856713913372\\
16.8736798456167	0.0384045362988417\\
17.7471566444319	0.0503502311150967\\
17.8245504708158	0.051469702076929\\
17.8936140131835	0.0524968303293107\\
17.9591099798286	0.0534693790249818\\
18.1970423011953	0.0570956742753097\\
18.5275101727965	0.0623472602171233\\
18.7257795996543	0.065579402560644\\
19.1301189491795	0.0724506737259603\\
20.419001900678	0.0965586114376649\\
20.5345493022396	0.0989388701595502\\
20.6255712260505	0.100819799693667\\
20.8032048820677	0.104572217973718\\
20.8700021627618	0.105994531719155\\
20.9520689538391	0.107768367312268\\
21.7564958792035	0.126317598441446\\
21.8817386642033	0.12931372108028\\
22.5437894476936	0.146328971641303\\
22.6817899913993	0.150120628197293\\
22.8056633094949	0.153535685811981\\
23.2089017386423	0.165171014716705\\
23.2788741791807	0.167237219719041\\
23.4011271722323	0.170886526890619\\
24.068155388261	0.191731787209768\\
24.2799374087356	0.198817541281819\\
25.0030735977424	0.223787688477636\\
25.0361000416575	0.224981948183427\\
25.2483211520811	0.232903507034806\\
25.2988885237791	0.234814529081305\\
25.4268059941836	0.239743505589374\\
25.4598584517404	0.24101727443319\\
25.7223898464206	0.251279263681333\\
25.8313799350982	0.255606094917766\\
26.4144886745996	0.27967789211273\\
26.7375946316465	0.293899996981189\\
26.767290837002	0.295215806834911\\
26.8463912234526	0.298787965065586\\
26.9813661455238	0.30498894026973\\
27.0947024012574	0.310281682147703\\
27.3027056190443	0.320131964523519\\
27.86108400019	0.347901895466748\\
27.9001447277918	0.349890656503511\\
28.0936273339478	0.359920608879591\\
28.6693237974641	0.390751125343691\\
28.7089037050449	0.392973545888132\\
29.1876406667765	0.420286483678346\\
29.2792524167075	0.425638673019499\\
29.6524377173031	0.447890463266923\\
29.6805639521756	0.449595705633428\\
29.9607223223047	0.466815408264239\\
29.9931131947895	0.46887276252771\\
30.0388038358801	0.471746302432507\\
30.3989599374939	0.494888772040639\\
30.4421660797242	0.497736939827192\\
30.967040352433	0.533291243166119\\
31.0331451334498	0.537920327136597\\
31.171454860039	0.547663073033732\\
31.2178316291706	0.550988885697379\\
31.710177542172	0.586925098420323\\
31.7436510530437	0.589424989629869\\
31.7578565135361	0.590508025049736\\
31.892976725222	0.600671918607276\\
32.0046482956358	0.609281917434089\\
32.0423136720226	0.612196318249964\\
32.0888408882051	0.615803879436029\\
32.2688754185423	0.629953998293246\\
32.4977720117533	0.648241837598107\\
32.546316924834	0.652140293070161\\
32.5713260860246	0.654175225461451\\
32.6419355370655	0.659905341354118\\
32.9000369334037	0.681305395489659\\
33.0233717508656	0.691750595032502\\
33.1066871183534	0.698897714227094\\
33.4127334607847	0.725823137161292\\
33.4853430198127	0.732311171330529\\
33.5055898186298	0.734110674066372\\
33.5652777854814	0.739491923332999\\
33.7068051385368	0.752221056055632\\
33.7438062625914	0.755562062487904\\
33.8389804051694	0.7642850557488\\
33.8777495306046	0.767837299816776\\
33.997570507962	0.778879776725567\\
34.0980684066128	0.788195474469887\\
34.1595025005458	0.793952732628988\\
34.6482717426961	0.840389904552538\\
34.6817510364766	0.843604787955137\\
34.7253613070762	0.847857420496823\\
34.8190025114968	0.857021623093239\\
34.9580136555824	0.870672707788252\\
34.9906271890306	0.873915913555415\\
35.0204122696415	0.876882238706486\\
35.0774793303551	0.882528125067388\\
35.1844026415771	0.893137505817968\\
35.2056828088377	0.89528956298782\\
35.2918297700291	0.903932014524386\\
35.3182092435169	0.906569534724319\\
35.5734015769425	0.932382272264832\\
35.8161333342341	0.957279548376142\\
35.9008611009443	0.966070839839105\\
36.0631866839154	0.982970440744623\\
36.0954702540553	0.986364109073527\\
36.1146072190962	0.988384880282666\\
36.1784138938476	0.995146402285251\\
36.224506869917	1\\
};
\addlegendentry{Upper LEO}

\addplot [color=black, dashed]
  table[row sep=crcr]{%
10	0\\
10	1\\
};
\end{axis}
\end{tikzpicture}%

%% file: main.bbl
\begin{thebibliography}{10}
\providecommand{\url}[1]{#1}
\csname url@samestyle\endcsname
\providecommand{\newblock}{\relax}
\providecommand{\bibinfo}[2]{#2}
\providecommand{\BIBentrySTDinterwordspacing}{\spaceskip=0pt\relax}
\providecommand{\BIBentryALTinterwordstretchfactor}{4}
\providecommand{\BIBentryALTinterwordspacing}{\spaceskip=\fontdimen2\font plus
\BIBentryALTinterwordstretchfactor\fontdimen3\font minus \fontdimen4\font\relax}
\providecommand{\BIBforeignlanguage}[2]{{%
\expandafter\ifx\csname l@#1\endcsname\relax
\typeout{** WARNING: IEEEtran.bst: No hyphenation pattern has been}%
\typeout{** loaded for the language `#1'. Using the pattern for}%
\typeout{** the default language instead.}%
\else
\language=\csname l@#1\endcsname
\fi
#2}}
\providecommand{\BIBdecl}{\relax}
\BIBdecl

\bibitem{abdelsadek2023future}
M.~Y. Abdelsadek, A.~U. Chaudhry, T.~Darwish, E.~Erdogan, G.~Karabulut-Kurt, P.~G. Madoery, O.~B. Yahia, and H.~Yanikomeroglu, ``Future space networks: Toward the next giant leap for humankind,'' \emph{IEEE Transactions on Communications}, vol.~71, no.~2, pp. 949--1007, 2023.

\bibitem{cheng2022service}
L.~Cheng, W.~Zhang, and Z.~Wang, ``{6G service-oriented space-air-ground integrated network: A survey},'' \emph{IEEE Access}, vol.~10, pp. 120\,000--120\,017, 2022.

\bibitem{ren2023review}
\BIBentryALTinterwordspacing
L.~Diana and P.~Dini, ``{Review on Hardware Devices and Software Techniques Enabling Neural Network Inference Onboard Satellites},'' \emph{Remote Sensing}, vol.~16, no.~21, 2024. [Online]. Available: \url{https://www.mdpi.com/2072-4292/16/21/3957}
\BIBentrySTDinterwordspacing

\bibitem{iqbal2023ai}
A.~Iqbal, M.-L. Tham, Y.~J. Wong, A.~Al-Habashna, G.~Wainer, Y.~X. Zhu, and T.~Dagiuklas, ``{Empowering Non-Terrestrial Networks With Artificial Intelligence: A Survey},'' \emph{IEEE Access}, vol.~11, 2023.

\bibitem{jia2020virtual}
Z.~Jia, M.~Sheng, J.~Li, Y.~Zhu, W.~Bai, and Z.~Han, ``{Virtual network functions orchestration in software defined LEO small satellite networks},'' in \emph{ICC 2020-2020 IEEE International Conference on Communications (ICC)}.\hskip 1em plus 0.5em minus 0.4em\relax IEEE, 2020, pp. 1--6.

\bibitem{3gpp_ntn}
M.~M. Saad, M.~A. Tariq, M.~T.~R. Khan, and D.~Kim, ``{Non-Terrestrial Networks: An Overview of 3GPP Release 17 \& 18},'' \emph{IEEE Internet of Things Magazine}, vol.~7, no.~1, pp. 20--26, 2024.

\bibitem{3GPP_TS28808}
\BIBentryALTinterwordspacing
{3GPP}, ``{Study on management and orchestration aspects of integrated satellite components in a 5G network},'' 3rd Generation Partnership Project (3GPP), Technical Report TR 28.808, April 2021, release 16. [Online]. Available: \url{https://www.3gpp.org/ftp/Specs/archive/28\_series/28.808/}
\BIBentrySTDinterwordspacing

\bibitem{mahboob2023revolutionizing}
S.~Mahboob and L.~Liu, ``{Revolutionizing Future Connectivity: A Contemporary Survey on AI-Empowered Satellite-Based Non-Terrestrial Networks in 6G},'' \emph{IEEE Communications Surveys \& Tutorials}, vol.~26, no.~2, pp. 1279--1321, 2024.

\bibitem{oranntn2025}
S.~Mahboob, J.~Dai, A.~Soysal, and L.~Liu, ``{Transforming Future 6G Networks via O-RAN-Empowered NTNs},'' \emph{IEEE Communications Magazine}, pp. 1--7, 2025.

\bibitem{wang2023satellite}
S.~Wang and Q.~Li, ``{Satellite Computing: Vision and challenges},'' \emph{IEEE Internet of Things Journal}, vol.~10, pp. 22\,514--22\,529, 2023.

\bibitem{chen2025spacegroundfluidai6g}
\BIBentryALTinterwordspacing
Q.~Chen, Z.~Wang, X.~Chen, J.~Wen, D.~Zhou, S.~Ji, M.~Sheng, and K.~Huang, ``{Space-ground Fluid AI for 6G Edge Intelligence},'' 2025. [Online]. Available: \url{https://arxiv.org/abs/2411.15845}
\BIBentrySTDinterwordspacing

\bibitem{DTNTN}
H.~Al-Hraishawi, M.~Alsenwi, J.~u. Rehman, E.~Lagunas, and S.~Chatzinotas, ``{Digital Twin for Enhanced Resource Allocation in 6G Non-Terrestrial Networks},'' \emph{IEEE Communications Magazine}, pp. 1--7, 2024.

\bibitem{clustering}
D.-H. Jung, G.~Im, J.-G. Ryu, S.~Park, H.~Yu, and J.~Choi, ``{Satellite Clustering for Non-Terrestrial Networks: Concept, Architectures, and Applications},'' \emph{IEEE Vehicular Technology Magazine}, vol.~18, no.~3, pp. 29--37, 2023.

\bibitem{dudukovich2024advances}
R.~Dudukovich, D.~Raible, B.~Tomko, N.~Kortas, E.~Schweinsberg, T.~Basciano, W.~Pohlchuck, J.~Deaton, J.~Nowakowski, and A.~Hylton, ``{Advances in high-rate delay tolerant networking on-board the international space station},'' in \emph{IEEE Space Mission Challenges for Information Technology-IEEE Space Computing Conference (IEEE SMC-IT/SCC)}, 2024.

\bibitem{Thz2}
W.~Jiang, Q.~Zhou, J.~He, M.~A. Habibi, S.~Melnyk, M.~El-Absi, B.~Han, M.~D. Renzo, H.~D. Schotten, F.-L. Luo, T.~S. El-Bawab, M.~Juntti, M.~Debbah, and V.~C.~M. Leung, ``Terahertz communications and sensing for 6g and beyond: A comprehensive review,'' \emph{IEEE Communications Surveys \& Tutorials}, vol.~26, no.~4, pp. 2326--2381, 2024.

\end{thebibliography}
